\documentclass[11pt,onecolumn,draftclsnofoot]{IEEEtran}
\usepackage{amssymb,amsmath,setspace,mathtools}
\usepackage{algorithmicx,algorithm,graphicx,subfigure,epstopdf}
\usepackage{cite,color,hyperref}
\usepackage{xfrac}
\usepackage{amsthm}
\usepackage{subfigure}
\usepackage{graphicx}
\usepackage{algpseudocode}

\DeclarePairedDelimiter{\ceil}{\lceil}{\rceil}
\newtheorem{Theo}{Theorem}
\newtheorem{remark}{Remark}
\newtheorem{corollary}{Corollary}

\begin{document}
\title{Jamming Bandits}
{ \author{\small{SaiDhiraj Amuru$^{\dagger}$, Cem~Tekin$^{\ddagger}$, Mihaela van der Schaar$^{\ddagger}$, R.~Michael Buehrer$^{\dagger}$\\}
{$^\dagger$Bradley Department of Electrical and Computer
Engineering, Virginia Tech\\
$^\ddagger$ Department of Electrical Engineering, UCLA}\\
Email: $\{$adhiraj,\ rbuehrer$\}$@vt.edu,\ cmtkn@ucla.edu,\ mihaela@ee.ucla.edu}

\date{}
\maketitle
\vspace{-45pt}
\noindent 
\begin{abstract}\vspace{-10pt}
Can an intelligent jammer learn and adapt to unknown environments in an electronic warfare-type scenario? In this paper, we answer this question in the positive, by developing
a cognitive jammer that adaptively and optimally disrupts the communication between a victim transmitter-receiver pair. We formalize the problem using a novel multi-armed bandit framework where the jammer can choose various physical layer parameters such as the signaling scheme, power level and the on-off/pulsing duration in an attempt to obtain power efficient jamming strategies. We first present novel online learning algorithms to maximize the jamming efficacy against static transmitter-receiver pairs and prove that our learning algorithm converges to the optimal (in terms of the error rate inflicted at the victim and the energy used) jamming strategy. Even more importantly, we prove that the rate of convergence to the optimal jamming strategy is sub-linear, i.e. the learning is fast in comparison to existing reinforcement learning algorithms, which is particularly important in dynamically changing wireless environments. Also, we characterize the performance of the proposed bandit-based learning algorithm against multiple static and adaptive transmitter-receiver pairs.
\end{abstract}

\vspace{-20pt}
\section{Introduction}
The inherent openness of the wireless medium makes it susceptible to adversarial attacks.
The vulnerabilities of a wireless system can be largely classified based on the capability of an adversary-
a) an eavesdropping attack in which the eavesdropper (passive adversary) can listen to the wireless channel and try to infer information (which if leaked may severely compromise data integrity) \cite{WynerWiretap}, \cite{CsiszarKorner}, b) a jamming attack, in which the jammer (active adversary) can transmit energy or information in order to disrupt reliable data transmission or reception  \cite{Basar-1983}-\hspace{-0.5pt}\cite{Kashyap-IT-2004} and c) a hybrid attack in which the adversary can either passively eavesdrop or actively jam any ongoing transmission \cite{Hybrid_Adversary_2}, \cite{Hybrid_Adversary_1}. In this paper, we study the ability of an agent to learn efficient jamming attacks against static and adaptive victim transmitter-receiver pairs.

Jamming has traditionally been studied by using either optimization or game-theoretic or information theoretic principles, see \cite{Azizoglu}-\hspace{-0.5pt}\cite{Shamai_Verdu} and references therein. The major disadvantage of these studies is that they assume the jammer has a lot of \emph{a priori} information about the strategies used by the (victim) transmitter-receiver pairs, channel gains, etc., which may not be available in practical scenarios. For instance, in our prior work \cite{GlobecomJamming}, we showed that it is not always optimal (in terms of the error rate) to match the jammer's signal to the victim's signaling scheme and that the optimal jamming signal follows a pulsed-jamming strategy. However, these optimal jamming strategies were obtained by assuming that the jammer has \emph{a priori} knowledge regarding the transmission strategy of the victim transmitter-receiver pair. In contrast to prior work (both ours and others), in this paper we develop online learning algorithms that learn the optimal jamming strategy by repeatedly interacting with the victim transmitter-receiver pair. Essentially, the jammer must learn to act in an unknown environment in order to maximize its total reward (e.g., jamming success rate). 

Numerous approaches have been proposed to learn how to act in unknown communication environments. A canonical example is reinforcement learning (RL) \cite{HarisVolos1}-\hspace{-0.5pt}\cite{Gulati2014}, in which a radio (agent) learns and adapts its transmission strategy using the transmission success feedback of the transmission actions it has used in the past. Specifically, it learns the optimal strategy by repeatedly interacting with the environment (for example, the wireless channel). During these interactions, the agent receives feedback indicating whether the actions performed were good or bad. The performance of the action taken is measured as a reward or cost, whose meaning and value depends on the specific application under consideration. For instance, the reward can be throughput, the negative of the energy cost, or a function of both these variables. In \cite{BeibeiWang}-\hspace{-0.5pt}\cite{CNSQLearning}, Q-Learning based algorithms were proposed to address jamming and anti-jamming strategies against adaptive opponents in multi-channel scenarios. It is well-known that such learning algorithms can guarantee optimality only asymptotically, for example as the number of packet transmissions goes to infinity. However, strategies with only asymptotic guarantees cannot be relied upon in mission-critical applications, where failure to achieve the required performance level will have severe consequences. For example, in jamming applications, the jammer needs to learn and adapt its strategy against its opponent in a timely manner. Hence, the rate of learning matters.

As discussed above, none of the previous works considered the learning performance of physical layer jamming strategies in electronic warfare environments where the jammer has limited to no knowledge about the victim transmitter-receiver pair. Further, the existing learning algorithms \cite{BeibeiWang}-\hspace{-0.5pt}\cite{Kleinberg} cannot be applied to the problem under consideration because a) none of the existing learning algorithms consider learning over a mixed (mixture of discrete/finite and continuous/infinite actions) action space and b) they do not give any performance guarantees for the jammer's actions. To fill this gap, in this paper, we present novel multi-armed bandit (MAB) algorithms to enable the jammer to learn the optimal physical layer jamming strategies, that were obtained in \cite{GlobecomJamming}, when the jammer has limited knowledge about the victim. While MAB algorithms have been used in the context of wireless communications to address the selection of a wireless channel in either cognitive radio networks \cite{CemMultiChannel}-\hspace{-0.5pt}\cite{Gai} or in the presence of an adversary \cite{QianWang}, or antenna selection in MIMO systems \cite{Gulati2014}, these works only consider learning over a finite action set. In contrast, the proposed algorithms in this paper enable the jammer to learn the optimal jamming strategies against both static and adaptive victim transmitter-receiver pairs by simultaneously choosing actions from both finite and infinite arm sets (i.e., they can either come from a continuous or a discrete space), that are defined based on the physical layer parameters of the jamming signal. In addition, our algorithms also provide time-dependent (not asymptotic) performance bounds on the jamming performance against static and adaptive victim transmitter-receiver pairs. We note that the algorithms proposed in this paper are novel even within the large area of multi-armed bandits. The major differences between our work and the prior work on multi-armed bandit problems (general works that are not related to jamming) are summarized in Table~\ref{tab1}. 

\begin{table}[h]
\caption{Comparison between related bandit works}
\label{tab1}
\centering
{\fontsize{10.5}{8.5}\selectfont
\begin{tabular}{|c|c|c|c|c|} \hline 
& Finite armed&Continuum armed &Adversarial&Our work \\ 
& bandits \cite{PeterAuer}&bandits\cite{Kleinberg}&bandits\cite{AuerExp3}& \\ \hline \hline
Regret bounds & Logarithmic & Sublinear & Sublinear & Sublinear  \\
(function of time)&&&&\\ \hline
Action rewards &i.i.d.&i.i.d.&adversarial &i.i.d. \\
&&&(worst-case)& \\ \hline
Action set & finite &continuous &finite &mixed \\ \hline
\end{tabular}
}
\vspace{-10pt}
\end{table}

We measure the jamming performance of a learning algorithm using the notion of \emph{regret}, which is defined as the difference between the cumulative reward of the optimal (for example, a strategy that minimizes the throughput of the victim while using minimum energy) jamming strategy when there is complete knowledge about the victim transmitter-receiver pair, and the cumulative reward achieved by the proposed learning algorithm. Any algorithm with regret that scales sub-linearly in time, will converge to the optimal strategy in terms of the average reward. These regret bounds can also provide a rate on how fast the jammer converges to the optimal strategy without having any \emph{a priori} knowledge about the victim's strategy and the wireless channel. As will be discussed in detail later, the feedback considered in this work is also minimal in comparison to the earlier jamming literature \cite{Azizoglu}-\hspace{-0.5pt}\cite{Shamai_Verdu}. 

The rest of the paper is organized as follows. We introduce the system model in Section~\ref{sec:SystemModel}. The jamming performance against static and adaptive transmitter-receiver pairs is considered in Sections~\ref{sec:FixedUser} and \ref{sec:StochasticUser} respectively, where we develop novel learning algorithms for the jammer and present high confidence bounds for its learning performance. Numerical results are presented in Section~\ref{Results} where we discuss the learning behavior in both single and multi-user scenarios and finally conclude the paper in Section~\ref{Conclusions}.

\section{System Model}\label{sec:SystemModel}
We first consider a single jammer and a single victim transmitter-receiver pair in a discrete time setting ($t=1,2,\ldots$). We assume that the data conveyed between the transmitter-receiver pair is mapped onto an unknown digital 
amplitude-phase constellation. The low pass equivalent of this signal
is represented as $x(t)=\sum_{m=-\infty}^{\infty}\sqrt{P_x}x_mg(t-mT)$, where $P_x$ is the average received signal 
power, $g(t)$ is the real valued pulse shape and $T$ is the symbol interval. The random variables 
$x_m$ denote the modulated symbols assumed to be uniformly distributed among all possible constellation points. Without loss of generality, the average energy of $g(t)$ and modulated symbols 
$E(|x_m|^2)$ are normalized to unity.\footnote{Any signal which follows a wireless standard (such as LTE) would have known parameters such as $g(t)$ and $T$ \cite{LTE}.}

\begin{figure}
\centering
\vspace{-20pt}
\includegraphics[width=0.45\textwidth]{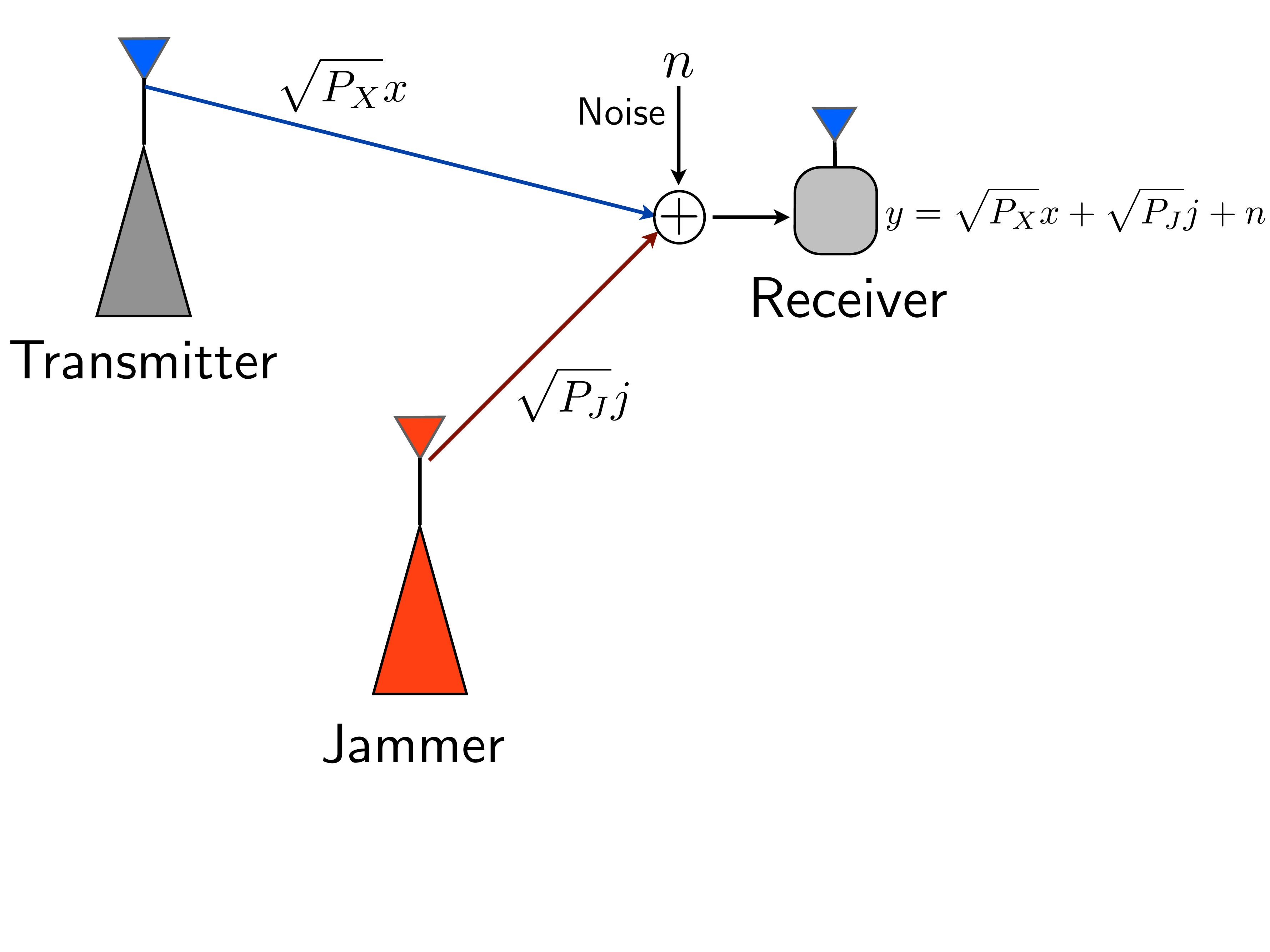}
\vspace{-40pt}
\caption{A wireless environment with victim transmitter-receiver pair and a jammer that intends to disrupt their communication.}
\vspace{-20pt}
\label{fig:SystemModel}
\end{figure}

It is assumed that $x(t)$ passes through an AWGN 
channel (received power is constant over the observation interval) 
while being attacked by a jamming signal represented as 
$j(t)=\sum_{m=-\infty}^{\infty}\sqrt{P_J}j_mg(t-mT)$, 
where $P_J$ is the average jamming signal power as seen at the victim receiver and 
$j_m$ denote the jamming signals with $E(|j_m|^2)\leq1$. 
Assuming a coherent receiver and perfect synchronization, the received signal 
after matched filtering and sampling at the symbol intervals is given by 
$y_k=y(t=kT)=\sqrt{P_x}x_k+\sqrt{P_J}j_k+n_k, \ k=1,2,..$ (as shown in Fig.~\ref{fig:SystemModel}),
where $n_k$ is the zero-mean additive white Gaussian noise with variance denoted by $\sigma^2$. Let $\mathrm{SNR}=\frac{P_x}{\sigma^2}$ and $\mathrm{JNR}=\frac{P_J}{\sigma^2}$. 
From \cite{GlobecomJamming}, the optimal jamming signal shares time between two different power levels one of which is $0$ and is hence defined by the on-off/pulsing duration $\rho$. In other words, the jammer sends the jamming signal $j(t)$ at power level $\mathrm{JNR}/\rho$ with probability $\rho$ and at power level $0$ (i.e., no jamming signal is sent) with probability $1-\rho$. For more details on the structure of the jamming signals, please see \cite{GlobecomJamming}. While the analysis shown in Sections~\ref{sec:FixedUser} and \ref{sec:StochasticUser} assumes coherent reception at the victim receiver (i.e., the jamming signal is coherently received along with the transmitter's signal), we consider the effects of a phase offset between these two signals in Section~\ref{Results}. The effects of a timing offset between $x$ and $j$ can also be addressed along similar lines, but is skipped in this paper due to a lack of space. 
\section{Jamming against a static transmitter-receiver pair} \label{sec:FixedUser}
In this section, we consider scenarios where the victim uses a fixed modulation scheme with a fixed $\mathrm{SNR}$. We propose an online learning algorithm for the jammer which learns the optimal power efficient jamming strategy over time, without knowing the victim's transmission strategy.

\subsection{Set of actions for the jammer}

At each time $t$ the jammer chooses its signaling scheme, power level and on-off/pulsing duration. A joint selection of these is also referred to as an action. We assume that the set of signaling schemes has $N_{mod}$ elements and the average power level belongs to the set  $\mathrm{JNR}\in[\mathrm{JNR}_{\min},\mathrm{JNR}_{\max}]$.\footnote{Although we use the variable $\mathrm{JNR}$ throughout this paper, it is crucial to notice that the proposed algorithms only need the knowledge of the power with which $j(t)$ is transmitted by the jammer and \textbf{do not} need to know the power of the jamming signal as seen at the victim receiver (which depends on the wireless channel whose knowledge is not available to the jammer). There is an unknown but consistent mapping between the jammer's transmit power and $\mathrm{JNR}$. The notation $\mathrm{JNR}$ is only used to make the exposition of the Theorems and the algorithms in this paper easier.} 
The jamming signal $j(t)$ is defined by the signaling scheme (for example AWGN, BPSK or QPSK ) and power level selected at time $t$. 
It is shown in \cite{GlobecomJamming} that the optimal jamming signal does not have a fixed power level, but instead it should alternate between two different power levels one of which is $0$. In other words, the jammer sends the jamming signal $j$ at power level $\mathrm{JNR}/\rho$ with probability $\rho$ and at $0$ (i.e., no jamming signal is sent) with probability $1-\rho$. Notice that such pulsed-jamming strategies enable the jammer to cause errors with a low average energy but a high instantaneous energy \cite{GlobecomJamming}. Therefore, the optimal jamming signal is characterized by the signaling scheme, the average power level and the pulse duration $\rho \in (0,1]$ which indicates the fraction of time that the jammer is transmitting. The jammer should learn these optimal physical layer parameters by first transmitting the jamming signal and then by observing the reward obtained for its actions. 

We formulate this learning problem as a \emph{mixed multi-armed bandit} (mixed-MAB) problem. In contrast to prior work on MAB problems, in a mixed-MAB the action space consists of both finite (signaling set) and continuum (power level, pulse duration) sets of actions. Next, we propose an online learning algorithm called {\em Jamming Bandits} (JB) where the jammer learns by repeatedly interacting with the transmitter-receiver pair. The jammer receives feedback about its actions by observing the acknowledgment /no acknowledgement (ACK/NACK) packets that are exchanged between the transmitter-receiver pair \cite{Patent_PER}. The average number of NACKs gives an estimate of the $PER$ which can be used to estimate the $SER$ as as $1-(1-PER)^{1/N_{sym}}$ where $N_{sym}$ is the number of symbols in one packet (other metrics such as throughput or goodput allowed can also be considered \cite{MilcomJamming}). Remember that the $SER$ and $PER$ are functions of the jammer's actions i.e., the signaling scheme, power level and pulse jamming ratio \cite{GlobecomJamming} and thereby allow the jammer to learn about its actions.  
\vspace{-11pt}
\subsection{MAB formulation}
The actions (also called the {\em arms}) of the mixed MAB are defined by the triplet $[$Signaling scheme, $\mathrm{JNR}$, $\rho$$]$. The strategy set $\mathcal{S}$, that constitutes $\mathrm{JNR}$ and $\rho$, is a compact subset of $({\mathbb{R}^+})^2$. For each time $t\in \{1,2,3,\ldots,n\} $, a cost (or objective) function (feedback metric)
$C_t : \{\boldsymbol{\mathcal{J}},\mathcal{S}\} \rightarrow \mathbb{R}$ is evaluated by the jammer, where $\boldsymbol{\mathcal{J}}$ indicates the set of signaling schemes. Since we are interested in finding power efficient jamming strategies that maximize the error rate at the victim receiver,
we define $C_t=\max(SER_t-SER_{target},0)/\mathrm{JNR}_t$ or $\max(PER_t-PER_{target},0)/\mathrm{JNR}_t$ where $\mathrm{JNR}_t$ indicates the average $\mathrm{JNR}$ used by the jammer at time $t$ and $SER_t$, $PER_t$ are the average symbol/packet error rate obtained by using a particular strategy $\{\mathcal{J}\in \boldsymbol{\mathcal{J}},\mathbf{s}\in\mathcal{S}\}$ at time $t$ and $SER_{target}$, $PER_{target}$ are the target error rates that should be achieved by the jammer (achieving a target $PER$ is a common constraint in practical wireless systems \cite{LTE} and this target is defined \emph{a priori}). The dependence of the cost function on the actions taken is unknown to the jammer \emph{a priori} because it is not aware of a) the victim's transmission strategy, b) the power of the signals $x$ and $j$ at the receiver (the probability of error is a function of these parameters as discussed in \cite{GlobecomJamming}) and hence needs to be learned over time in order to optimize the jamming strategy. The jammer does this by trying to maximize $C_t$ as it intends to maximize the error rate at the victim receiver using minimum energy. 

When the action set is a continuum of arms, most existing MAB works \cite{Kleinberg} assume that the 
arms that are close to each other (in terms of the Euclidean distance), yield similar expected costs. Such assumptions on the cost function will at least help in learning strategies that are close to the optimal strategy (in terms of the achievable cost function) if not the optimal strategy \cite{Kleinberg}. In this paper, for the first time in a wireless communication setting, we prove that this condition indeed holds true i.e., it is not an assumption but rather an intrinsic (proven) feature of our problem and we show how to evaluate the H\"{o}lder continuity parameters for these cost functions. Specifically, Theorem~\ref{Lipschitz_Theorem} shows that this similarity condition indeed holds true when the cost function is $SER$ and extends it to other commonly used cost functions in wireless scenarios. The result in this Theorem is crucial for deriving the regret and high confidence bounds of the proposed learning algorithm. 

Formally, the expected or average cost function $\bar{C}(\mathcal{J},\mathbf{s}) : \{\boldsymbol{\mathcal{J}},\mathcal{S}\}\hspace{-1pt}\rightarrow \hspace{-1pt}\mathbb{R}$ is shown to be uniformly locally H\"{o}lder continuous with constant $L\in[0,\infty)$,
exponent $\alpha\in (0,1]$ and restriction $\delta>0$. More specifically, the uniformly locally H\"{o}lder continuity condition (described with respect to the continuous arm parameters) is given by,\vspace{-10pt}
\begin{align}\label{generic_cost_function_Lipschitz}
|\bar{C}(\mathcal{J},\mathbf{s})-\bar{C}(\mathcal{J},\mathbf{s}')|\leq L||\mathbf{s}-\mathbf{s}'||^{\alpha},
\end{align}
for all $\mathbf{s},\mathbf{s}'\in  \mathcal{S}$ with $0\leq||\mathbf{s}-\mathbf{s}'||\leq \delta$ \cite{Audibert} ($||\mathbf{s}||$ denotes the Euclidean norm of the continuous $2\times 1$ action vector $\mathbf{s}$). The best strategy $\mathbf{s}^*$ satisfies $\arg \min_{\mathbf{s}\in\mathcal{S}}\bar{C}(\mathcal{J},\mathbf{s})$ for a signaling scheme $\mathcal{J}$.
As we will shown next, the algorithms proposed in this paper only require the jammer to know a bound on $L$ and $\alpha$, since it is not always possible to be aware of the cost function (its dependence on the actions taken) \emph{a priori}. 

\begin{Theo}\label{Lipschitz_Theorem}
For any set of strategies used by the victim and the jammer, the resultant $SER$ is uniformly locally H\"{o}lder continuous. 
\vspace{-8pt}\end{Theo}
\textit{Proof}: See Appendix A. In an online setting, the H\"{o}lder continuity parameters $L$ and $\alpha$ can be estimated if the jammer has knowledge about the victim's transmission strategy, else a bound on $L$ and $\alpha$ works. 

We now give an illustrative example for Theorem~\ref{Lipschitz_Theorem}. Consider the scenario where both the jammer and the victim use BPSK modulated signals. The average $SER$ (first we show for the case when $\rho=1$ which will be used to prove the result for $\rho\in(0,1]$) is given by \cite{GlobecomJamming}
\vspace{-5pt}
{\small\begin{align}
p_e(\mathrm{SNR},\mathrm{JNR})=\frac{1}{4}\left(erfc\left(\frac{\sqrt{\mathrm{SNR}}+\sqrt{\mathrm{JNR}}}{\sqrt{2}}\right)+erfc\left(\frac{\sqrt{\mathrm{SNR}}-\sqrt{\mathrm{JNR}}}{\sqrt{2}}\right)\right),
\end{align}}
where $erfc$ is the complementary error function. To show the H\"{o}lder continuity of the above expression, consider $\mathrm{JNR}_1$ and $\mathrm{JNR}_2$ such that $|\mathrm{JNR}_1-\mathrm{JNR}_2|\leq \delta$, for some $\delta>0$ (i.e., to consider the case of local H\"{o}lder continuity). Then by using the Taylor series expansion of the $erfc$ function and ignoring the higher order terms i.e., $erfc(x)\approx1-\frac{2}{\sqrt{\pi}}x+\frac{2}{3\sqrt{\pi}}x^3$,
we have
\vspace{-10pt}
{\small \begin{align}\label{BPSK_Lip}
p_e(\mathrm{SNR},\hspace{-1pt}\mathrm{JNR}_1)\hspace{-2pt}-\hspace{-2pt}p_e(\mathrm{SNR},\hspace{-2pt}\mathrm{JNR}_2)&\hspace{-2pt}\approx \hspace{-2pt}\sqrt{\frac{\mathrm{SNR}}{8\pi}}\hspace{-2pt}\left(\mathrm{JNR}_1\hspace{-2pt}-\hspace{-2pt}\mathrm{JNR}_2\right)\leq \sqrt{\frac{\mathrm{SNR}_{\max}}{8\pi}}\hspace{-2pt}\left(\mathrm{JNR}_1\hspace{-2pt}-\hspace{-2pt}\mathrm{JNR}_2\right),
\end{align}}
where $\mathrm{SNR}_{\max}$ relates to the maximum received power level of the victim signal (practical wireless communication devices have limitations on the maximum power levels that can be used). This shows that $SER$ satisfies the H\"{o}lder continuity property when $\rho=1$.

For the case of a pulsed jamming signal i.e., $\rho\in(0,1]$, the $SER$ is given by
$\rho p_e(\mathrm{SNR},\mathrm{JNR}/\rho)+(1-\rho) p_e(\mathrm{SNR},0)$. The second term is obviously H\"{o}lder continuous with respect to the strategy vector $\mathbf{s}=\{\mathrm{JNR},\rho\}$ for $L_1=1, \alpha_1=1$. For the first term, consider the probability of error at the strategies $\mathbf{s}_1=\{\mathrm{JNR}_1,\rho_1\}$ and $\mathbf{s}_2=\{\mathrm{JNR}_2,\rho_2\}$.  To prove the H\"{o}lder continuity, we consider the expression $\rho_1p_e(\mathrm{SNR},\mathrm{JNR}_1/\rho_1)-\rho_2p_e(\mathrm{SNR},\mathrm{JNR}_2/\rho_2)= \Big\{\rho_1p_e\Big(\mathrm{SNR},\frac{\mathrm{JNR}_1}{\rho_1}\Big)-\rho_1p_e\Big(\mathrm{SNR},\frac{\mathrm{JNR}_2}{\rho_1}\Big)\Big\}+\Big\{\rho_1p_e\Big(\mathrm{SNR},\frac{\mathrm{JNR}_2}{\rho_1}\Big)-\rho_2p_e\Big(\mathrm{SNR},\frac{\mathrm{JNR}_2}{\rho_2}\Big)\Big\}$. Again, the first term in this expression is H\"{o}lder continuous with $L_2=\sqrt{\frac{\mathrm{SNR}_{\max}}{8\pi}},\alpha_2=1$ which follows from \eqref{BPSK_Lip}. Using the Taylor series for $erfc$ and after some manipulations, the second term in this expression can be written as
{\small \begin{align}
&\rho_1p_e(\mathrm{SNR},\frac{\mathrm{JNR}_2}{\rho_1})-\rho_2p_e(\mathrm{SNR},\frac{\mathrm{JNR}_2}{\rho_2})\leq (\rho_1-\rho_2)\frac{erfc(\mathrm{SNR})}{2}\nonumber \\
&\leq \frac{erfc(\mathrm{SNR})}{2}\sqrt{(\mathrm{JNR}_1-\mathrm{JNR}_2)^2+(\rho_1-\rho_2)^2}\triangleq L_3||\mathbf{s}-\mathbf{s}'||^{\alpha_3}. 
\end{align} \vspace{-5pt}}
Overall, with $L=3\min(L_1,L_2,L_3)$ and $\alpha=1$, the $SER$ obtained under pulsed jamming is also H\"{o}lder continuous. In general, since the jammer does not know the victim signals' parameters, it is not aware of the exact structure of the $SER$ expression and hence it can use the worst case $L$ and $\alpha$ (across all possible scenarios that may occur in a real time scenario) to account for the H\"{o}lder continuity of $C_t$. 

\begin{corollary}
$PER$ and $\max(PER-PER_{target},0)/\mathrm{JNR}$ are H\"{o}lder continuous.
\end{corollary}\vspace{-10pt}
\emph{Proof:} 
$PER$ can be expressed in terms of the $SER$. For example, $PER=1-(1-SER)^{N_{sym}}$ when a packet is said to be in error if at least one symbol in the packet is received incorrectly. Since Theorem~\ref{Lipschitz_Theorem} shows that $SER$ is H\"{o}lder continuous, it follows that $PER$ and as a consequence $\max(PER-PER_{target},0)/\mathrm{JNR}$ are also H\"{o}lder continuous (remember that $\mathrm{JNR}\in[\mathrm{JNR}_{\min},\mathrm{JNR}_{\max}]$). It is worth noticing that the H\"{o}lder continuity parameters $L$ and $\alpha$ depend on the physical layer signaling parameters such as a) the modulation schemes used by the victim and the jammer and b) $\mathrm{SNR}$ of the victim signal.


\vspace{-10pt}

\subsection{Proposed Algorithm}

The proposed Jamming Bandits (JB) algorithm is shown in Algorithm~\ref{alg:1}. At each time $t$, JB forms an estimate $\hat{C}_t$ on the cost function $\bar{C}$, which is an average of the costs observed over the first $t-1$ time slots. Since some dimensions of the joint action set are continuous, and have infinitely many elements, it is not possible to learn the cost function for each of these values, because it will require a certain amount of time to explore each action from these infinite sets, which thereby cannot be completed in finite time. To overcome this, JB discretizes them and then approximately learns the cost function among these discretized versions. 
For example, $\rho$ is discretized as $\{1/M,2/M,\ldots,1\}$ and $\mathrm{JNR}$ is discretized as $\mathrm{JNR}_{\min}+(\mathrm{JNR}_{\max}-\mathrm{JNR}_{\min})*\{1/M,2/M,\ldots,1\}$, where $M$ is the {\em discretization} parameter. The performance of JB will depend on $M$, hence, we will also compute the optimal value of $M$ in the following sections.  

JB divides the entire time horizon $n$ into several rounds with different durations. Within every round (the duration $T$ of each round is also adaptive as shown in Alg.~\ref{alg:1}), JB uses a different discretization parameter $M$ to create the discretized joint action set, and learns the best jamming strategy over this set, as shown in Fig.~\ref{fig:JBRound}. The discretization $M$ increases with the number of rounds. Its value given in line 2 of Algorithm~\ref{alg:1} balances the loss incurred due to exploring actions in the discretized set and the loss incurred due to the sub-optimality resulting from the discretization. The various losses incurred and the derivation of the optimal value for $M$ will be explained in detail in Theorem~\ref{theorem1}. 

\begin{algorithm}[h]
\caption{Jamming Bandits (JB)}
\begin{algorithmic}[1]
\Statex T$\leftarrow$ 1 
\While{$T\leq n$} \\
\hspace{15pt}	$M\leftarrow \ceil{(\sqrt{\frac{T}{\mathrm{log} T}}L2^{\alpha/2})^{\frac{1}{1+\alpha}}}$ \\
\hspace{15pt}	Initialize UCB1 algorithm \cite{PeterAuer} with strategy set $\{$AWGN,BPSK,QPSK$\}$$\times$ $\{1/M,2/M,\ldots,1\}$$\times$$\mathrm{JNR}_{\min}$+$(\mathrm{JNR}_{\max}$-$\mathrm{JNR}_{\min})$*$\{1/M,2/M,\ldots,1\}$, where $\times$ indicates the Cartesian product.
	\For {$t=T,T+1,\ldots,\min(2T-1,n)$} \\
\hspace{25pt}   Choose arm $\{\mathcal{J}_t,\mathbf{s}_t\}$ from UCB1\cite{PeterAuer} \\
\hspace{25pt}	Play $\{\mathcal{J}_t,\mathbf{s}_t\}$ and then estimate $C_t(\mathcal{J}_t,\mathbf{s}_t)$ using the ACK/NACK packets\\
\hspace{25pt}	For each arm in the strategy set, update its index using $C_t(\mathcal{J}_t,\mathbf{s}_t)$.
	\EndFor \\
	$T\leftarrow 2T$
\EndWhile
\end{algorithmic}\label{alg:1}
\end{algorithm}
\begin{figure}
\vspace{-15pt}
\centering
\includegraphics[width=0.6\columnwidth]{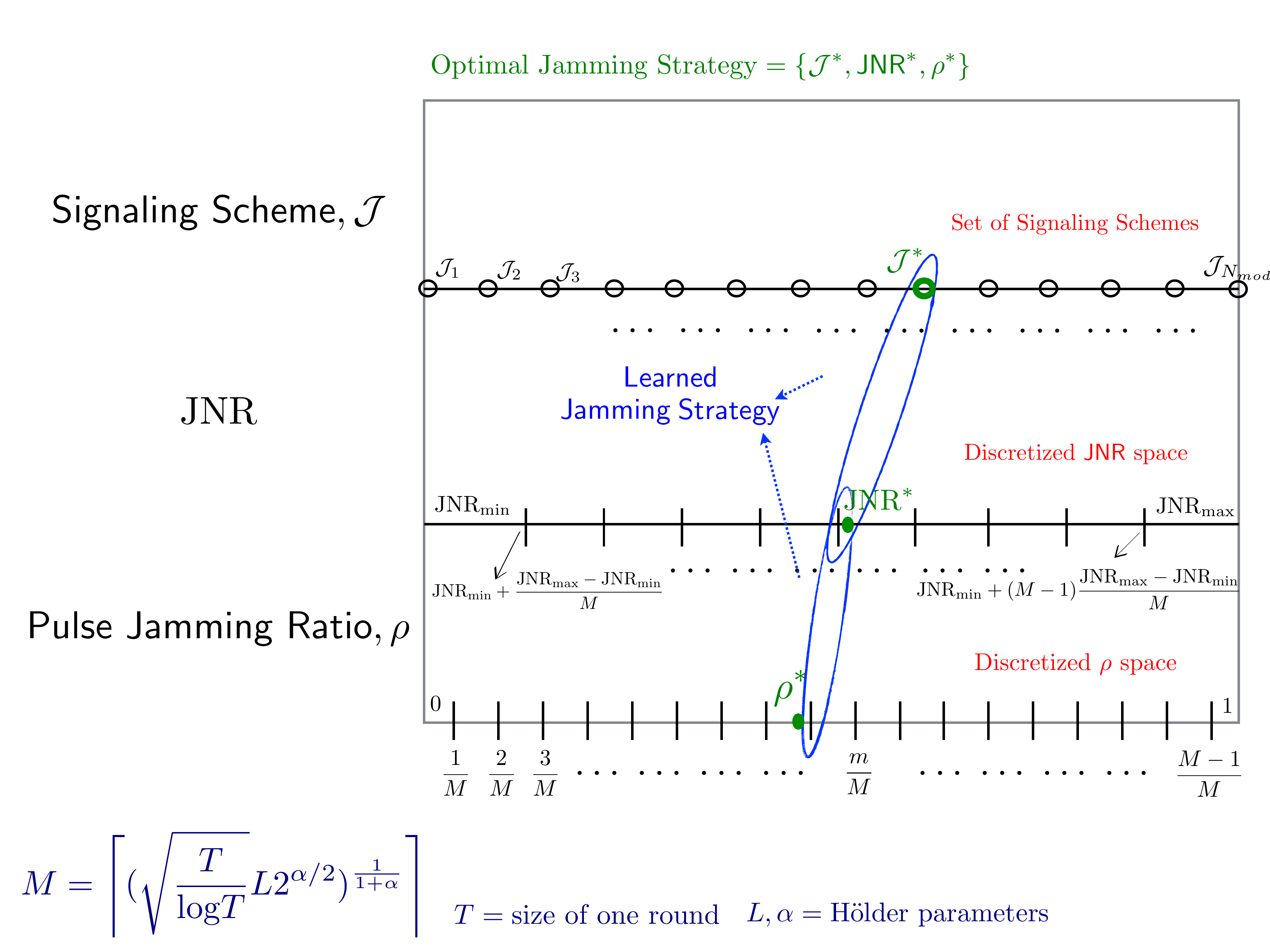}
\vspace{-15pt}
\caption{An illustration of learning in one round of JB. It is possible that the optimal strategy denoted by $\{\mathcal{J}^*,\mathsf{JNR}^*,\rho^*\}$ lies out of the set of discretized strategies. In such a case the jammer learns the best discretized strategy, but based on the value of the discretization parameter $M$, the loss incurred by using this strategy with respect to the optimal strategy can be bounded using the H\"{o}lder continuity condition. The value of the discretization $M$ is shown in the figure and Alg.~\ref{alg:1}.}
\label{fig:JBRound}
\vspace{-25pt}
\end{figure}\vspace{-15pt}
Another advantage of JB is that the jammer does not need to know the time horizon $n$. Time horizon $n$ is only given as an input to JB to indicate the stopping time. All our results in this paper hold true for any time horizon $n$. This is achieved by increasing the time duration of the inner loop in JB to $2T$ at the end of every round (popularly known as the doubling trick \cite{Kleinberg}). The inner loop can use any of the standard finite-armed MAB algorithms such as UCB1 \cite{PeterAuer}, which is shown in Algorithm~\ref{alg:ucb} for completeness.

\begin{algorithm}[h]
\caption{Upper confidence bound-based MAB algorithm - UCB1}
\begin{algorithmic}[1]
\Statex \textbf{Initialization}: Play each arm once
\Statex \textbf{Loop}: \newline \hspace{15pt} Use signaling scheme $\mathcal{J}$, power $\mathrm{JNR}$, pulse jamming ratio $\rho$, which maximizes $\hat{C}(\mathcal{J},\underbrace{\mathrm{JNR},\rho}_{\mathbf{s}})+\sqrt{\frac{2\mathrm{log}t}{u_{\mathcal{J},\mathbf{s}}}}$ where $t$ is the time duration since the start of the algorithm, $u_{\mathcal{J},\mathbf{s}}$ is the number of times the arm $\{\mathcal{J},\mathbf{s}\}$ has been played and $\hat{C}(\mathcal{J},\underbrace{\mathrm{JNR},\rho}_{\mathbf{s}})$ is the estimated average reward obtained from this arm. 
\end{algorithmic}
\label{alg:ucb}
\end{algorithm}\vspace{-22pt}
\vspace{-22pt}
\subsection{Upper bound on the regret}
For the proposed algorithm, the $n$-step regret $R_n$ is the expected difference in the total cost between the strategies chosen by the proposed algorithm i.e., $\{\mathcal{J}_1,\mathbf{s}_1\},\{\mathcal{J}_1,\mathbf{s}_2\},\ldots,\{\mathcal{J}_n,\mathbf{s}_n\}$ and the best strategy $\{\mathcal{J}^*,\mathbf{s}^*\}$. More specifically, we have
$R_n=\mathbf{E}\Big[\sum_{t=1}^n\Big(C_t(\mathcal{J}^*,\mathbf{s}^*)-C_t(\mathcal{J}_t,\mathbf{s}_t)\Big)\Big]$,
 where the expectation is over all the possible strategies that can be chosen by the proposed algorithm. 
Here we present an upper bound on the cumulative regret that is incurred by the jammer when it uses Algorithm~\ref{alg:1} to minimize regret or in other words maximize the cost/objective function. 
\begin{Theo}\label{theorem1}
The regret of JB is $\mathcal{O}(N_{mod}{n^{\frac{\alpha+2}{2(\alpha+1)}} (\mathrm{log} n)^{\frac{\alpha}{2(\alpha+1)}}})$.
\end{Theo}\vspace{-11pt}
\textit{Proof}: See Appendix B.
\vspace{-11pt}
\begin{remark}
The upper bound on regret increases as $N_{mod}$ increases. This is because the jammer now has to spend more time in identifying the optimal jamming signaling scheme. This does not mean that the jammer is doing worse, since as $N_{mod}$ increases, the jamming performance of the {\em benchmark} against which the regret is calculated also gets better. Hence, the jammer will converge to a better strategy, though it learns more slowly.  Further, the regret decreases as $\alpha$ increases because higher values of $\alpha$ indicate that it is easier to separate strategies that are close (in Euclidean distance) to each other.
\end{remark}\vspace{-10pt}

\begin{corollary}
The average cumulative regret of JB converges to $0$. Its convergence rate is given as $\mathcal{O}(n^{\frac{-\alpha}{2(\alpha+1)}}(\mathrm{log}\hspace{1.25pt}n)^{\frac{\alpha}{2(\alpha+1)}})$.
\end{corollary}\vspace{-10pt}
The average cumulative regret converges to $0$ as $n$ increases. These results establish the learning performance i.e., the rate of learning (how fast the regret converges to $0$) of JB and indicate the speed at which the jammer learns the optimal jamming strategy using Algorithm~\ref{alg:1}. Since the proposed algorithms and hence their regret bounds are dependent only on $L$ and $\alpha$, which are in turn a function of the various signal parameters such as the modulation schemes used by the victim and the jammer, the wireless channel model i.e., AWGN channel, Rayleigh fading channel etc, the proposed algorithms can be extended to a wide variety of wireless scenarios by only changing these parameters. The exact values of $L$ and $\alpha$
need not be known in these cases (because the jammer may not have complete knowledge of the wireless channel conditions), the worst case $L$ and $\alpha$ (as shown in the BPSK example below Theorem~\ref{Lipschitz_Theorem}) can be used in the proposed JB algorithm. 
\vspace{-10pt}
\subsection{High Confidence Bounds}
The confidence bounds provide an \emph{a priori} probabilistic guarantee on the desired level of jamming performance (e.g., $SER$ or $PER$) that can be achieved at a given time. We first present the one-step confidence bounds i.e., the instantaneous regret and later show the confidence level obtained on the cumulative regret over $n$ time steps.

The sub-optimality gap $\Delta_i$ of the $i$th arm $\{\mathcal{J}^i,\mathbf{s}^i\}$ (recall that $N_{mod}M^2$ arms can be chosen in one round of JB), is defined as $\bar{C}(\mathcal{J}^*,\mathbf{s}^*)-\bar{C}(\mathcal{J}^i,\mathbf{s}^i)$. We say that an arm is sub-optimal if it belongs to the set $U_{>}$ (set of arms whose sub-optimality gap exceeds a threshold based on the required jamming confidence level) which is defined in detail in Appendix C.
Let $u_i(t)$ denote the total number of times the $i$th arm has been chosen until time $t$ and
$U(T)$ indicate the set of time instants $t\in[1,T]$ for which $u_i(t) \leq \frac{8\log(T)}{\Delta_i^2}$ for some sub-optimal arm $i \in {\cal U}_>$.

\begin{Theo}(i)\label{TheoremOneStepRegret}
 Let $\delta = 2 \times 2^{\frac{3\alpha+2}{2(1+\alpha)}}L^{\frac{1}{1+\alpha}}\left(\frac{\mathrm{log}T}{T}\right)^{\frac{\alpha}{2(1+\alpha)}}$ and $M$ be defined as in Algorithm~\ref{alg:1}. Then for any $t\in[1,T]\backslash U(T)$, with probability at least $1 - 2(N_{mod}+M^2)t^{-4}$, the expected cost of the chosen jamming strategy $(\mathcal{J}_t,\mathbf{s}_t)$ is at most $\bar{C}(\mathcal{J}^*,\mathbf{s}^*) + \delta$. 
In other words, \newline$P\left(\bar{C}(\mathcal{J}^*,\mathbf{s}^*)-\bar{C}(\mathcal{J}_t,\mathbf{s}_t) >\delta \right)\leq2(N_{mod}+M^2)t^{-4}$. \newline
(ii) We also have \vspace{-10pt}
\begin{align}
E[|U(T)|] &\leq \sum_{t=1}^T P(\textrm{a sub-optimal arm } i \in {\cal U}_> \textrm{ is chosen at } t) \notag \\
& \leq 8 \sum_{i \in {\cal U}_>} \left( \frac{\log T}{\Delta_i^2}  \right) + \left(1 + \frac{\pi^2}{3} \right) |U_>|, \notag
\end{align}
which means that our confidence bounds hold in all except logarithmically many time slots in expectation.
\end{Theo}
\emph{Proof:} See Appendix C. 
\begin{remark}
A lower bound on the sub-optimality gap i.e., $\Delta_{\min}=\min_{i\in  {\cal U}_>}\Delta_i$, can be used to approximately estimate $U(T)$. For instance, in a wireless setting when $SER$ is used as the cost function, if the jammer is aware of the smallest tolerable error in $SER$ that is allowed, then it can approximately evaluate $U(T)$. A detailed discussion on how the jammer can estimate $U(T)$ is given in Appendix C.
\end{remark}

\begin{corollary}
The one-step regret converges to zero in probability i.e.,
\begin{align}
\lim_{T \rightarrow \infty}\left( \lim_{t \rightarrow T} P\left(\bar{C}(\mathcal{J}^*,\mathbf{s}^*)-\bar{C}(\mathcal{J}_t,\mathbf{s}_t) >\delta \right) \right)=0.    \notag
\end{align}
\end{corollary}

Theorem~\ref{TheoremOneStepRegret} can be used to achieve desired confidence levels about the jamming performance, which is particularly important in military settings. In order to achieve a desired confidence level (e.g., about the $SER$ inflicted at the victim receiver) $\delta$ at each time step, the probability of choosing a jamming action that incurs regret more than $\delta$ must be very small. In order to achieve this objective, the jammer can set $M$ as $\max \{ (\frac{2^{\frac{\alpha+4}{2}}L}{\delta})^{1/\alpha},  \ceil{(\sqrt{\frac{T}{\mathrm{log} T}}L2^{\alpha/2})^{\frac{1}{1+\alpha}}} \}$. By doing this, the jammer will not only guarantee a small regret at every time step, but also chooses an arm that is within $\delta$ of the optimal arm at every time step with high probability. Hence, the one time step confidence about the jamming performance can be translated into overall jamming confidence. It was, however, observed that the proposed algorithm performs significantly better than predicted by this bound (Section~\ref{Results}). 


\begin{Theo}\label{TheoremCumulativeRegret}
For any signaling scheme $\mathcal{J}$ chosen by the jammer, 
$P\Big(\sum_{t=1}^T(\bar{C}(\mathcal{J},\mathbf{s}^*)-\bar{C}(\mathcal{J},\mathbf{s}_t))$
$>\Big(\frac{8}{3\epsilon}\Big(\frac{T}{\mathrm{log}T}\Big)^{\frac{4}{1+\alpha}}\Big)^{1/3}\Big)$ $<\epsilon$, $\forall\ \epsilon>0$.
\end{Theo}\vspace{-10pt}
\emph{Proof:} See Appendix C. Using Theorem~\ref{TheoremCumulativeRegret}, a confidence bound on the overall cumulative regret defined as $\sum_{t=1}^T[\bar{C}(\mathcal{J}^*,\mathbf{s}^*)-\bar{C}(\mathcal{J}_t,\mathbf{s}_t)]$ can be directly obtained as shown in Appendix C. This bound indicates the overall confidence acquired by the jammer. The regret performance of JB will be discussed in more detail via numerical results in Section~\ref{Results}. 

\begin{Theo}\label{TheoremEstimateProb}
Let  $\delta= 2 \times 2^{\frac{5\alpha+4}{2(1+\alpha)}}L^{\frac{1}{1+\alpha}}\left(\frac{\mathrm{log}T}{T}\right)^{\frac{\alpha}{2(1+\alpha)}}$ and $M$ be defined as in JB. Then, for any $t \in[1,T]\backslash U(T)$, the jammer knows that with probability at least $1-2(N_{mod}+M^2)t^{-4}-t^{-16}$, the true expected cost of the optimal strategy is at most $\hat{C}(\mathcal{J}_t,\mathbf{s}_t) + \delta$, where $\hat{C}(\mathcal{J}_t,\mathbf{s}_t)$ is the sample mean estimate of $\bar{C}(\mathcal{J}_t,\mathbf{s}_t)$, the expected reward of strategy $(\mathcal{J}_t,\mathbf{s}_t)$ selected by the jammer at time $t$.
\end{Theo}\vspace{-0pt}
\emph{Proof:} See Appendix D. Theorem~\ref{TheoremEstimateProb} presents a high confidence bound on the estimated cost function of any strategy used by the jammer. Such high confidence bounds (Theorems 3-5) will enable the jammer to make decisions on the jamming duration and jamming budget, which is explained below with an example.
Again, this is a worst case bound and the proposed algorithm performs much better than predicted by the bound as will be discussed in detail in Section~\ref{Results}.

\begin{figure}
\centering
\includegraphics[width=0.85\columnwidth]{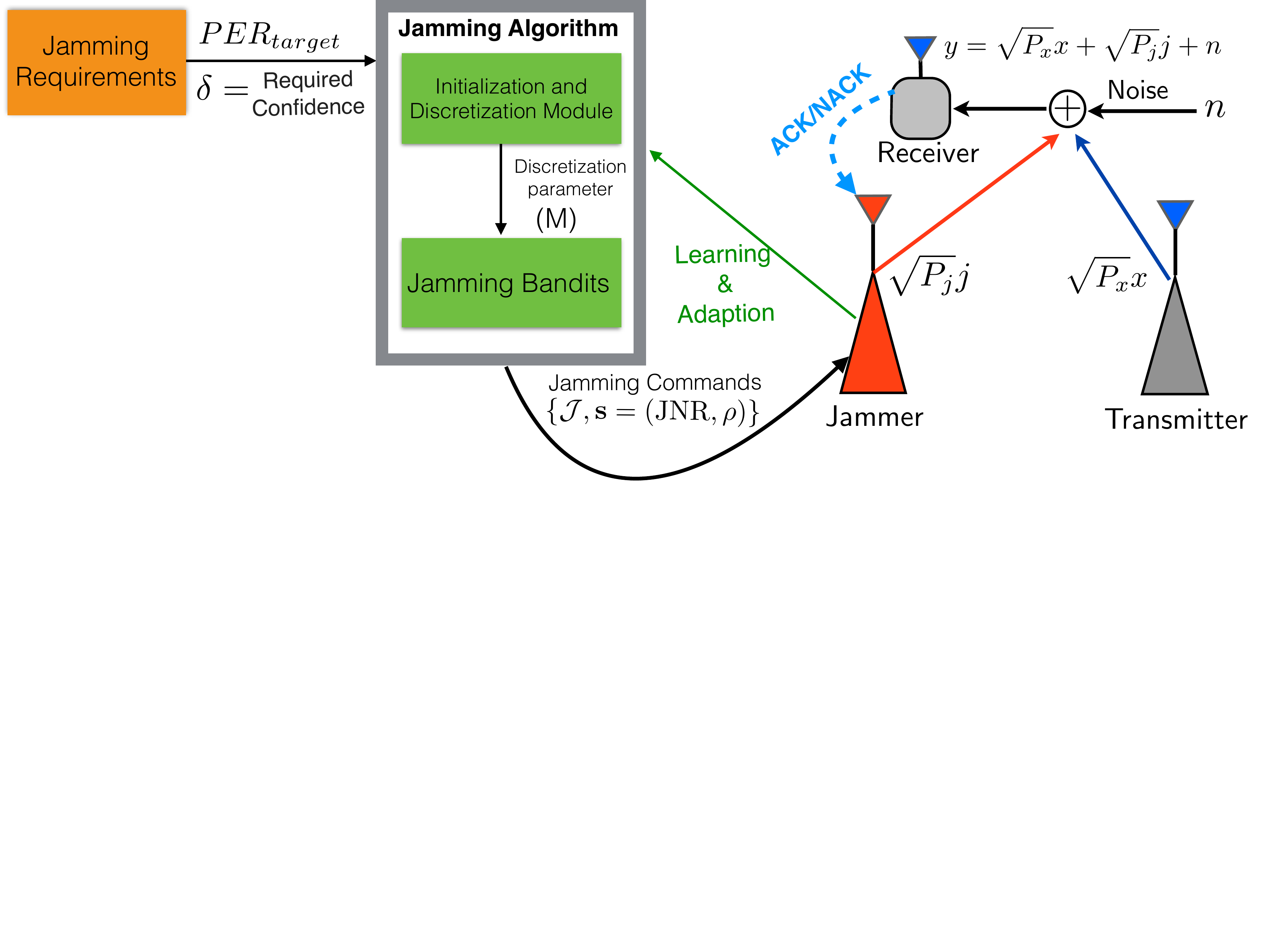}
\vspace{-160pt}
\caption{Using Theorems~\ref{TheoremOneStepRegret} and \ref{TheoremEstimateProb} in a real time jamming environment.}
\label{fig:Thm3_4_Fig}
\vspace{-25pt}
\end{figure}

\begin{remark}
Fig.~\ref{fig:Thm3_4_Fig} summarizes the importance and usability of Theorems~\ref{TheoremOneStepRegret} and \ref{TheoremEstimateProb} in realtime wireless communication environments. The high confidence bounds for the regret help the jammer decide the number of symbols (or packets) to be jammed to disrupt the communication between the victim transmitter-receiver pair. For example, such confidence is necessary in scenarios where the victim uses erasure or rateless codes and/or HARQ-based transmission schemes. In the case of rateless codes, a message of length $N$ is encoded into an infinitely long new message sequence of length $\hat{N}>>N$ (for example, by using random linear combinations) out of which any $N$ are linearly independent. Upon successfully receiving $N$ such messages, the entire message can be recovered. Under such scenarios, the high confidence bounds help the jammer to decide the number of packets/ time instants to jam successfully in order to disrupt the wireless link between the transmitter-receiver pair. 

For instance, when $M=15$, we have at large time $t$, $\delta>0.01$, i.e., $P(SER^*-\hat{SER}_t>0.01)=0$, where $SER^*$ is the optimal average $SER$ achievable and $\hat{SER}_t$ is the estimated $SER$ achieved by the strategy used at time $t$. If the jammer estimates $SER$ as $0.065$ then the best estimate of the $SER^*$ indicates that it is less than or equal to $0.075$. Using such knowledge, the jammer can identify the minimum number of packets it has to jam so as to disrupt the communication and prevent the exchange of a certain number of packets (which in applications such as video transmission can completely break down the system). As an example, consider the case when packets of length $100$ symbols are exchanged and that a packet is said to be in error only when there are more than $10$ errors in the packet. Thus, in order to jam 100 packets successfully the jammer needs to affect at least $463$ packets on an average if $SER^*$ (which corresponds to $PER=0.2167$) was achievable. However, since it can only achieve $\hat{SER}=0.065$ i.e., $\hat{PER}=0.1153$, it has to jam at least $865$ packets on an average to have sufficient confidence regarding its jamming performance. The jammer can accordingly plan its energy budget/ jamming duration etc. by using such knowledge.
\end{remark}
\vspace{-20pt}
\subsection{Improving convergence via arm elimination}
When the number of signaling schemes that the jammer can choose from is large or when $\alpha$ is small (i.e., it is difficult to separate the arms that are close to each other), then the learning speed using JB can be relatively slow. We now present an algorithm to improve the learning rate and convergence speed of JB under such scenarios. In order to achieve this, Algorithm~\ref{alg:1} is modified to use the UCB-Improved algorithm \cite{AuerElimination} inside the inner loop of JB instead of UCB1. The UCB-Improved algorithm eliminates sub-optimal arms (that are evaluated in terms of the mean rewards and the confidence intervals), in order to avoid exploring the sub-optimal arms (which is important in electronic warfare scenarios). The modified algorithm and the associated UCB-Improved algorithm are shown in Algorithms~\ref{alg:2} and \ref{alg:3} respectively.
\begin{algorithm}[h]
\caption{Jamming Bandits with Arm Elimination}
\begin{algorithmic}[1]
\Statex T$\leftarrow$ 1
\While{$T\leq n$} \\
\hspace{15pt}	Initialize UCB-Improved \cite{AuerElimination} algorithm with the strategy set 
$\{$AWGN,BPSK,QPSK$\}$$\times$ $\{1/M,2/M,\ldots,1\}$$\times$$\mathrm{JNR}_{\min}$+$(\mathrm{JNR}_{\max}$-$\mathrm{JNR}_{\min})$*$\{1/M,2/M,\ldots,1\}$, where $\times$ indicates the Cartesian product.
	\For {$t=T,T+1,\ldots,\min(2T-1,n)$} \\
	\hspace{25pt} Use the UCB-Improved \cite{AuerElimination} MAB Algorithm to eliminate sub-optimal arms
	\EndFor \\
	$T\leftarrow 2T$
\EndWhile
\end{algorithmic}\label{alg:2}
\end{algorithm}
\vspace{-10pt}
\begin{algorithm}[h]
\caption{UCB-Improved}
\begin{algorithmic}[1]
\Statex Input the set of arms $A$ and time horizon $T$
\Statex $\tilde{\Delta}_0=0,B_0=A$
\For {rounds $m=0,1,2,\ldots,\frac{1}{2}\mathrm{log}_2\frac{T}{e}$} \\
\hspace{15pt} \textbf{Arm Selection} \\
\hspace{15pt} If $|B_m|>1$, choose each arm in $B_m$ for $n_m=\ceil{\frac{2\mathrm{log}(T\tilde{\Delta}^2_m)}
{\tilde{\Delta}_m^2}}$ \\
\hspace{15pt} Else choose the remaining arm until time $T$ \\
\hspace{15pt} \textbf{Arm Elimination} \\
\hspace{15pt} Delete arm $i$ in the set $B_m$ for which\ 
$\left(\bar{C}_i+\sqrt{\frac{\mathrm{log}(T\tilde{\Delta}^2_m))}{2n_m}}\right)$$<$$\max_{j\in B_m}\left(\bar{C}_j-\sqrt{\frac{\mathrm{log}(T\tilde{\Delta}^2_m))}{2n_m}}\right)$
\hspace{15pt} to obtain the set of new arms $B_{m+1}$; $\bar{C}_i$ is the average cost incurred by playing arm $i$. \\
\hspace{15pt} \textbf{Reset} $\tilde{\Delta}_m$ : $\tilde{\Delta}_{m+1}=\tilde{\Delta}_m/2$ .
\EndFor
\end{algorithmic}\label{alg:3}
\end{algorithm}

To obtain the value of $M$ i.e., the discretization for $\mathrm{JNR}$ and $\rho$, we used numerical optimization tools to solve $TL\left(\frac{2}{M^2}\right)^{\frac{\alpha}{2}}-\left(\sqrt{M^2T}\frac{\mathrm{log}(M^2\mathrm{log}(M^2))}{\sqrt{\mathrm{log}(M^2)}}\right)=0$.
See Appendix E for more details. Later in Section~\ref{Results}, we show the benefits of using this algorithm via numerical simulations. The regret bounds can be derived along similar lines to Theorems~1-5 by using the properties of the UCB-Improved algorithm \cite{AuerElimination}.

\section{Learning jamming strategies against a time-varying user}\label{sec:StochasticUser}
In this section, we consider scenarios where the victim transmitter-receiver pair can choose their strategies in a time-varying manner.\footnote{The model considered in this formulation is different from the \emph{adversarial scenarios} studied in the context of MAB algorithms \cite{AuerExp3}. In the adversarial bandit cases, the adversary (or the victim in this current context) observes the action of the jammer and then assigns a reward function either based on the jammers' current action or on the entire history of jammers' actions. However, in the current scenario we assume that the user picks a strategy in an i.i.d manner independent of the jammer. Considering learning algorithms in adversarial scenarios is reserved for future work.} We specifically consider two scenarios a) when the victim changes its strategies in an i.i.d. fashion and b) when the victim is adapting its transmission strategies to overcome the interference seen in the wireless channel.\footnote{While the victim is not entirely adaptive against the jammers' strategies, it is adaptive in the sense that it can choose from a set of strategies to overcome the jamming/ interference effects. For example, it can be adaptive based on the $PER$ seen at the victim receiver. This scenario is discussed in detail in Section~\ref{Results}.}
The worst case jammer's performance can be understood by considering a victim that changes its strategies in an i.i.d. fashion. For example, such i.i.d. strategies are commonly employed in a multichannel wireless system where the victim can randomly hop onto different channels (either in a pre-defined or an un-coordinated fashion \cite{Strasser}) to probabilistically avoid jamming/ interference. 
The randomized strategies chosen by the victim can confuse the jammer regarding its performance. For instance, if the jammer continues using the same strategy irrespective of the victim's strategy, then the jammers' performance will be easily degraded. However, if the jammer is capable of anticipating such random changes by the victim and learns the jamming strategies, then it can disrupt the communication irrespective of the victims' strategies.

We assume that the victim can modify its power levels and the modulation scheme to adapt to the wireless environment (the most widely used adaption strategy \cite{Poisel}). Again we allow the jammer to learn the optimal jamming strategy by optimizing the $3$ actions, namely signaling scheme, $\mathrm{JNR}$ and $\rho$ as before. The jammer has to learn its actions without any knowledge regarding the victim's strategy set and any possible distribution that the victim may employ to choose from this strategy set. We use Algorithm~\ref{alg:1} and not Algorithm~\ref{alg:2} to address such dynamic scenarios because eliminating arms in such a time-varying environment may not always be beneficial. For example, a certain arm might not be good against one strategy used by the victim but might be the optimal strategy when the victim changes its  strategy. 

While the regret bounds presented below assume that the victim employs a random unknown distribution over its strategy set and chooses its actions in an i.i.d. manner (also referred to as stochastic strategies) i.e., scenario (a) mentioned earlier, we discuss the jammer's performance against any strategy (i.e., without any predefined distribution over the strategies, for example, increase the power levels when the $PER$ increases) employed by the victim (which includes scenario (b)) in Section~\ref{Results}.

\subsubsection{Upper bound on the regret}
Let $\{p_i\}_{i=1}^{|\mathcal{P}|}$ denote the probability distribution with which the victim selects its strategies in an i.i.d manner, from a set consisting of $|\mathcal{P}|$ number of possible strategies. The jammer is not aware of this distribution chosen by the victim and needs to learn the optimal strategy by repeatedly interacting with the victim. The regret under such scenarios is defined as
$R_n=\mathbf{E}\Big[\sum_{t=1}^n\Big(C_t(\mathcal{J}^*,\mathbf{s}^*)-C_t(\mathcal{J}_t,\mathbf{s}_t)\Big)\Big]$, 
where the expectation is over the random strategies chosen by the jammer as well as the victim (which is different from the formulation in Section~\ref{sec:FixedUser}).
Thus, the above expression can be re-written as
$R_n=\mathbf{E}\Big[\sum_{t=1}^n\sum_{i=1}^{|\mathcal{P}|}p_i\Big(C^{i}_t(\mathcal{J}^*,\mathbf{s}^*)-C^{i}_t(\mathcal{J}_t,\mathbf{s}_t)\Big)\Big]$, with $C_t^i$ indicating the cost function when the victim uses strategy $i$ with probability $p_i$ and the expectation is now taken only over the strategies chosen by the jammer. 

\begin{Theo}\label{TheoStochasticRegret}
The regret of JB when the victim employs stochastic strategies is $\mathcal{O}(N_{mod}n^{\frac{\alpha+2}{2(\alpha+1)}} (\mathrm{log}n)^{\frac{\alpha}{2(\alpha+1)}} )$.
\end{Theo}\vspace{-11pt}
\emph{Proof:} See Appendix F. This is an upper bound on the cumulative regret incurred by JB under such stochastic scenarios. Similar to the regret incurred by JB in Theorem 1, the regret under stochastic cases also converges to $0$ as $\mathcal{O}(n^{\frac{-\alpha}{2(\alpha+1)}}(\mathrm{log}\hspace{1.25pt}n)^{\frac{\alpha}{2(\alpha+1)}})$.
The one step confidence bounds similar to Theorems 3-5 can be derived even in this case but are skipped due to lack of space. 
\begin{remark}
When the victim is adapting its strategies based on the error rates observed over a given time duration (as is typically done in practical wireless communication systems), we show that by employing sliding-window based algorithms, the jammer can effectively track the changes in the victim and jam it in a power efficient manner. This is discussed more in detail in the next section.
\end{remark}

\vspace{-11pt}
\section{Numerical Results}\label{Results}
We first discuss the learning behavior of the jammer against a transmitter-receiver pair that employs a static strategy and later consider the performance against adaptive strategies. To validate the learning performance, we compare the results against the optimal jamming signals that are obtained when the jammer has complete knowledge about the victim \cite{GlobecomJamming}. It is assumed that the victim and the jammer send $1$ packet with $10000$ symbols at any time $t$. 
A packet is said to be in error if at least $10\%$ of the symbols are received in error at the victim receiver so as to capture the effect of error correction coding schemes. 
The minimum and the maximum $\mathrm{SNR}, \mathrm{JNR}$ levels are taken to be $0\ dB$ and $20\ dB$ respectively. The set of signaling schemes for the transmitter-receiver pair is $\{BPSK, QPSK\}$ and for the jammer is $\{AWGN, BPSK, QPSK\}$\footnote{It is very easy to extend the results in this paper and \cite{GlobecomJamming} to PAM and QAM signals of any constellation size.} \cite{GlobecomJamming} i.e., $N_{mod}=3$. \

\subsection{Fixed user strategy}
The jammer uses $SER$ or $PER$ inflicted at the victim receiver (estimated using the ACK and NACK packets) as feedback to learn the optimal jamming strategy. We first consider a scenario where the $\mathrm{JNR}$ is fixed and the jammer can optimize its jamming strategy by choosing the optimal signaling scheme $\mathcal{J}^*$ and the associated pulse jamming ratio $\rho^*$. These results enable comparison with previously known results obtained via an optimization framework with full knowledge about the victim as discussed in \cite{GlobecomJamming}. Note that unlike \cite{GlobecomJamming}, the jammer here does not know the signaling  parameters of the victim signal, and hence it cannot solve an optimization problem to find the optimal jamming strategy. In contrast, it learns over time the optimal strategy by simply learning the expected reward of each strategy it tries.
%
%

\begin{figure}

\begin{minipage}[t]{0.5\textwidth}
\centering
\includegraphics[width=2.5in]{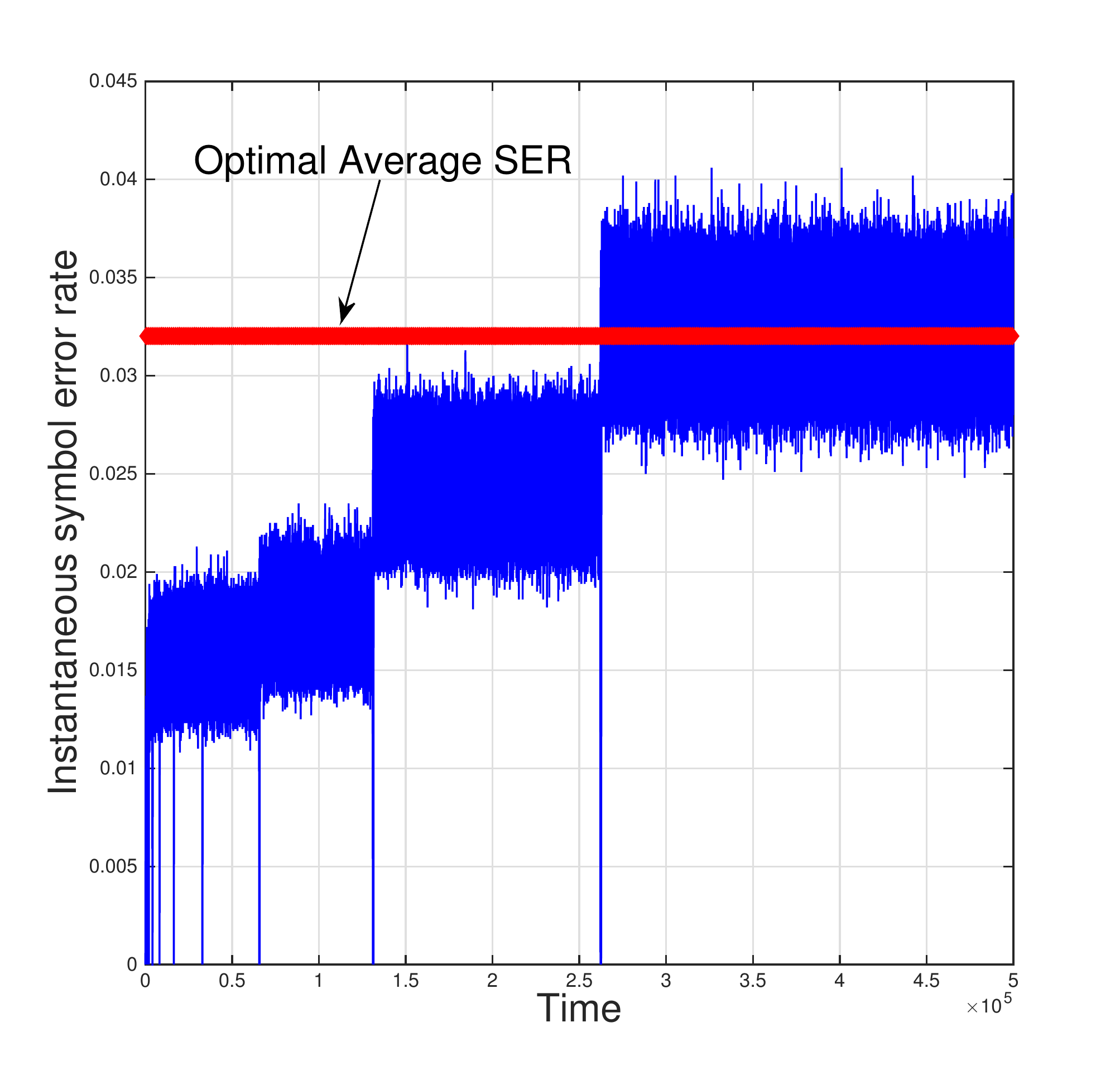}
\vspace{-15pt}
\caption{Instantaneous SER achieved by the JB algorithm \newline when $\mathrm{JNR}=10dB$, $\mathrm{SNR}=20dB$ and the victim uses \newline BPSK.}
\label{figinstantaneous}
\end{minipage}
\begin{minipage}[t]{0.5\textwidth}
\centering
\includegraphics[width=2.5in]{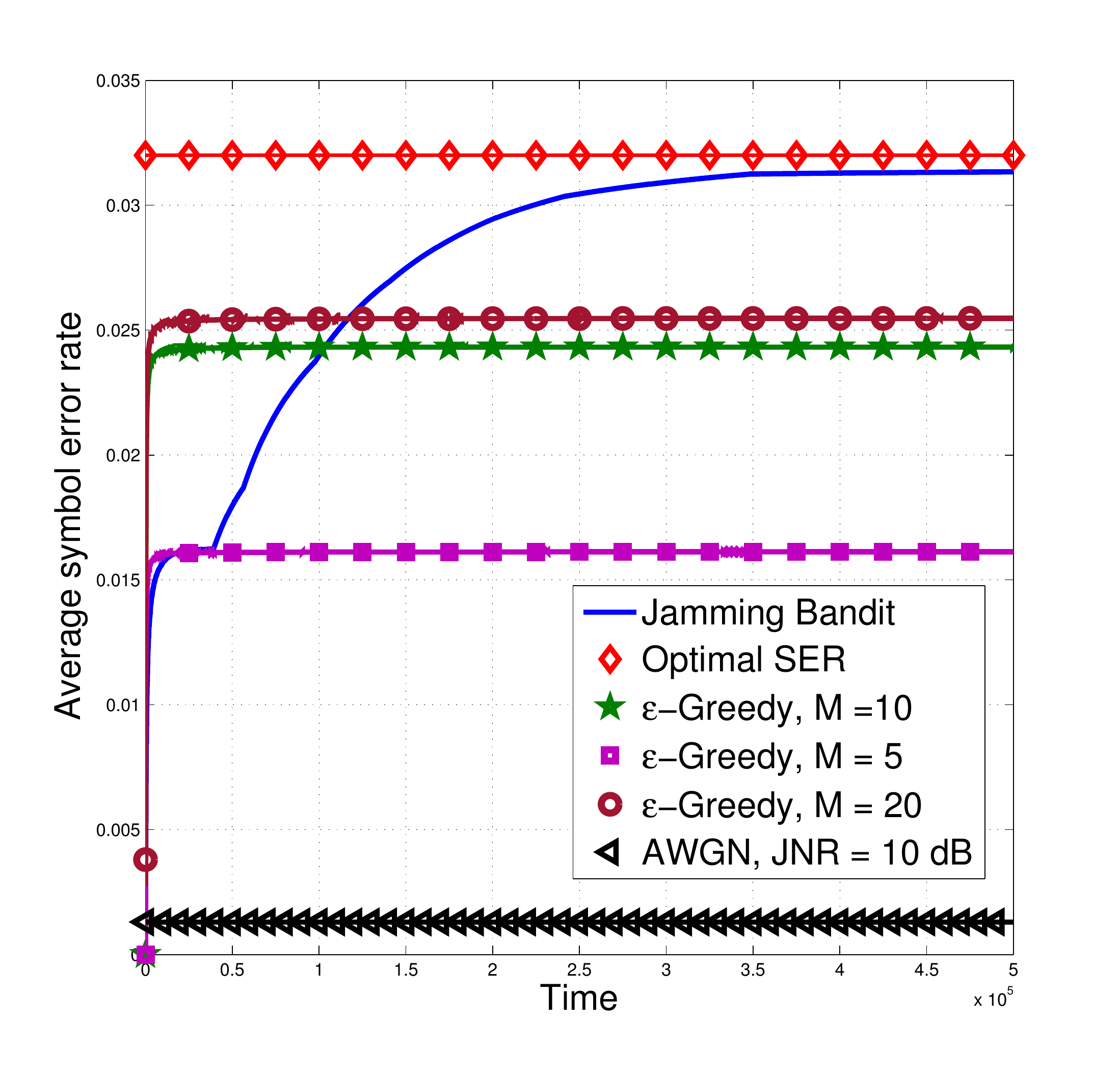}
\vspace{-15pt}
\caption{Average SER achieved by the jammer when $\mathrm{JNR}=10dB$, $\mathrm{SNR}=20dB$ and the victim uses BPSK. The jammer learns to use BPSK with $\rho=0.078$ using JB. The learning performance of the $\epsilon$-greedy learning algorithm with various discretization factors $M$ is also shown.}
\label{fig1}
\end{minipage}
\vspace{-20pt}
\end{figure}

Figs.~\ref{figinstantaneous}-\ref{fig_non_coherent} show the results obtained in this setting (fixed $\mathrm{SNR}$, modulation scheme for the victim and fixed $\mathrm{JNR}$). For a fair comparison with \cite{GlobecomJamming}, we initially assume that the jammer can directly estimate the $SER$ inflicted at the victim receiver. We will shortly discuss the more practical setting in which the jammer can only estimate $PER$. In all these figures, it is seen that the jammers' performance converges to that of the optimal jamming strategies\cite{GlobecomJamming}. For example, in Figs.~\ref{figinstantaneous} and \ref{fig1}, when the victim transmitter-receiver pair exchange a BPSK modulated signal at $\mathrm{SNR}=20$ dB, the jammer learns to use BPSK signaling at $\mathrm{JNR}=10$ dB and $\rho=0.078$ which is in agreement with the results presented in \cite{GlobecomJamming}. 

\begin{figure}
\begin{minipage}[t]{0.5\textwidth}
\centering
\includegraphics[width=2.5in]{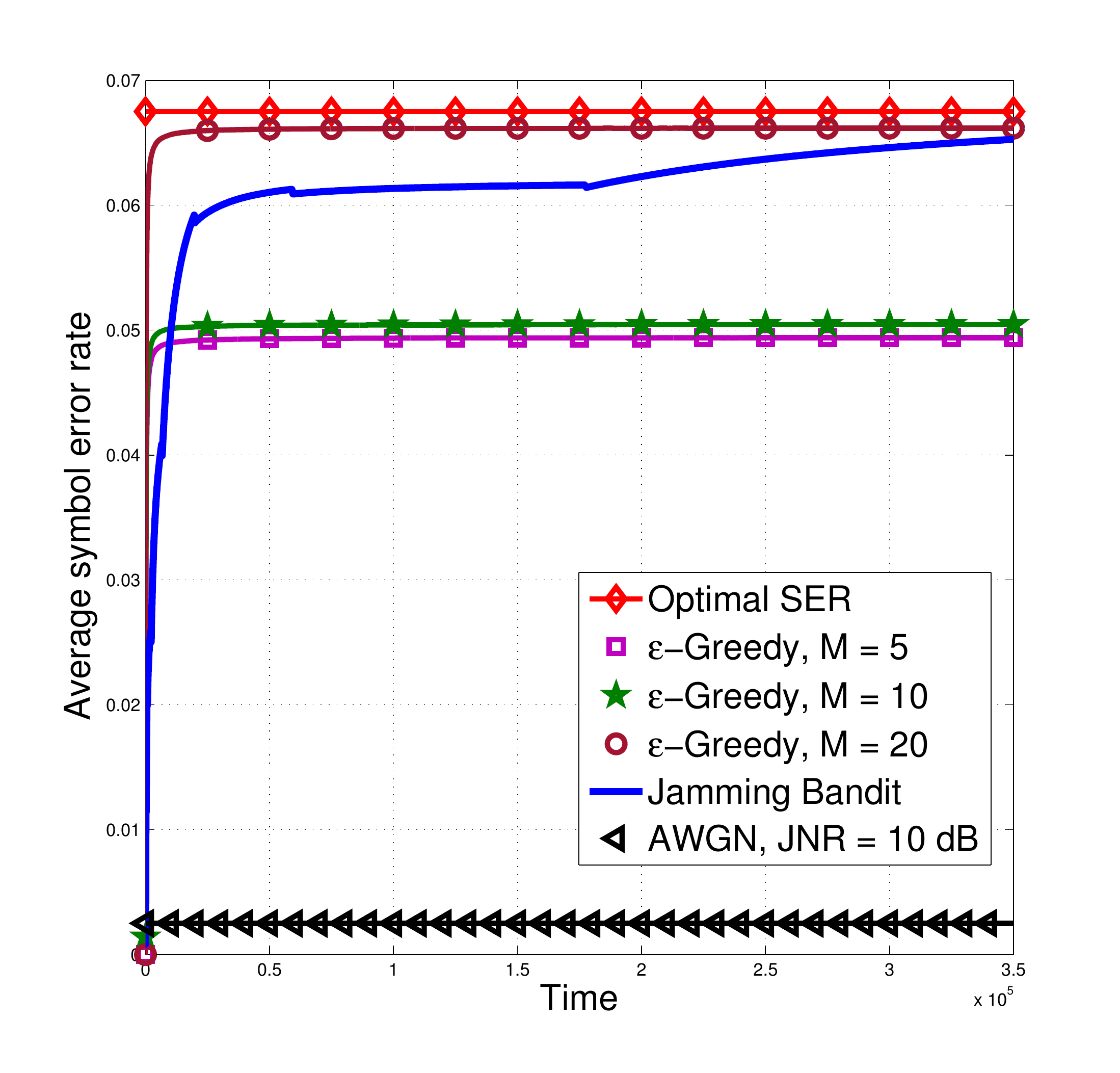}
\vspace{-20pt}
\caption{Learning the optimal jamming strategy when $\mathrm{JNR}=10dB$, $\mathrm{SNR}=20dB$ and the victim uses QPSK modulation scheme. The jammer learns to use QPSK signaling scheme with $\rho=0.087$.}
\vspace{-20pt}
\label{fig2}
\end{minipage}\hspace{5pt}
\begin{minipage}[t]{0.5\textwidth}
\includegraphics[width=2.5in]{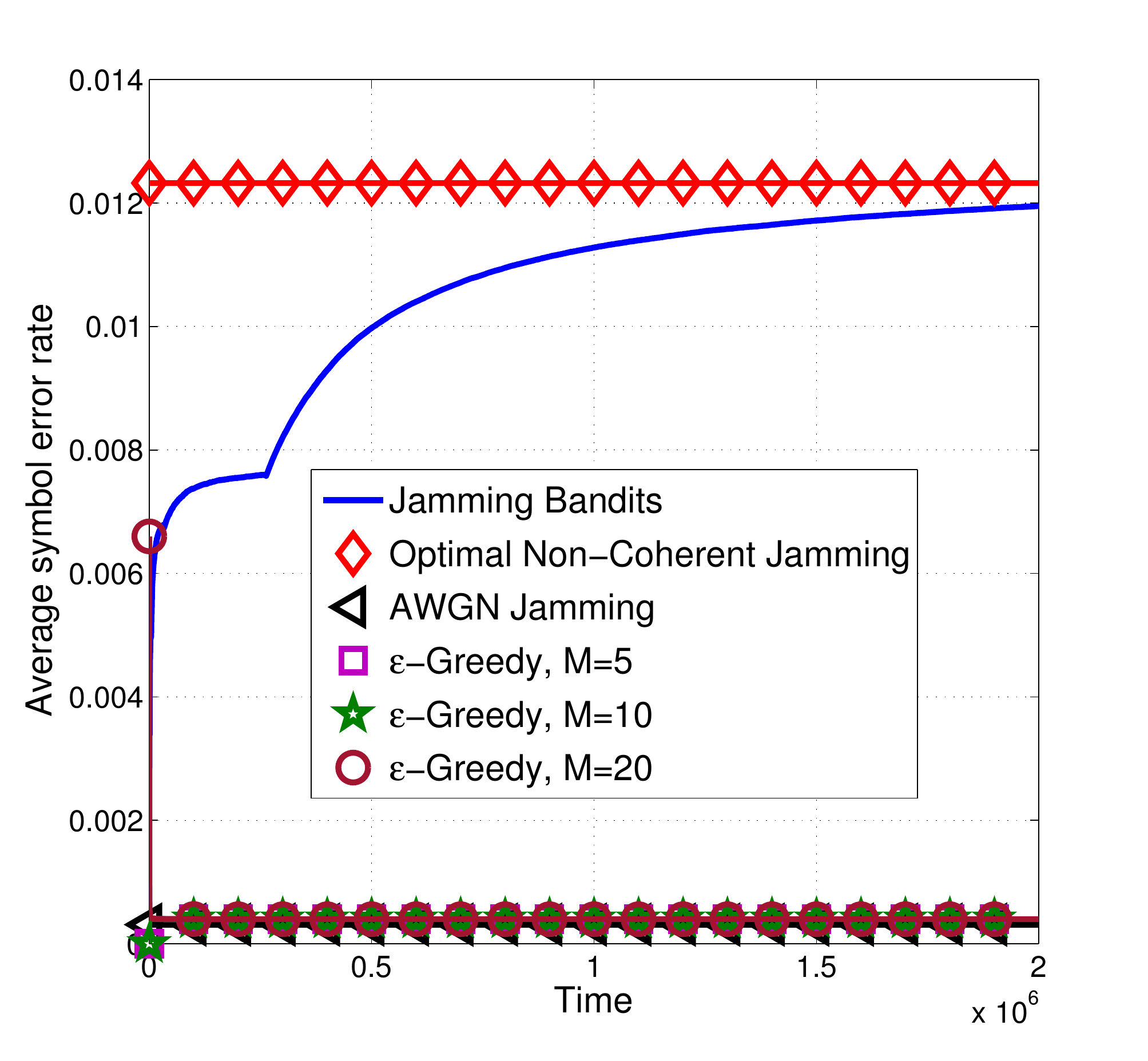}
\vspace{-20pt}
\caption{Average SER achieved by the jammer when $\mathrm{JNR}=10dB$, $\mathrm{SNR}=20dB$ and the victim uses BPSK and there is a phase offset between the two signals. The jammer learns to use BPSK with $\rho=0.051$ using JB. The learning performance of the $\epsilon$-greedy learning algorithm with various discretization factors $M$ is also shown.}
\label{fig_non_coherent}
\end{minipage}
\vspace{-20pt}
\end{figure}

Fig.~\ref{figinstantaneous} shows the instantaneous learning performance of the jammer in terms of the $SER$ achieved by using the JB algorithm. The variation in the achieved $SER$ after convergence is only due to the wireless channel. The time instants at which the $SER$ varies a lot, i.e., the dips in $SER$ seen in these results are due to the exploration phases performed when a new value of discretization i.e., $M$ is chosen by the algorithm (recall from Algorithm~\ref{alg:1} that for every round the discretization $M$ is re-evaluated). Fig.~\ref{fig1} shows the average SER attained by this learning algorithm. Also shown in Fig.~\ref{fig1} is the performance of the the $\epsilon$-greedy learning algorithm \cite{PeterAuer} with exponentially decreasing exploration probability $\epsilon(t)=\epsilon^{\frac{t}{10}}$ (initial exploration probability is taken to be $0.9$) and various discretization factors $M$. In the $\epsilon$-greedy learning algorithm, the jammer explores (i.e., it tries new strategies) with probability $\epsilon(t)$ and exploits (i.e., uses the best known strategy that has been tried thus far) with probability $1-\epsilon(t)$. It is seen that unless the optimal discretization factor $M$ is known (so that the optimal strategy is one among the possible strategies that can be chosen by the $\epsilon$-Greedy algorithm), the $\epsilon$-greedy algorithm performs significantly worse in comparison to the novel bandit-based learning algorithm. 

Similar results were observed in the case of QPSK signaling as seen in Fig.~\ref{fig2}. Notice that while the $\epsilon$-greedy algorithm with discretization $M=20$ did not achieve satisfactory results in the BPSK signaling scenario, it achieved close to optimal results in the QPSK signaling scenario as seen in Fig.~\ref{fig2}. Thus, the performance of the $\epsilon$-greedy algorithm highly depends on $M$, and it can be sub-optimal if $M$ is chosen incorrectly. However, in our learning setting it is not possible to know the optimal $M$ \emph{a priori}. Also, the performance of AWGN jamming (which is the most widely used jamming signal \cite{Stark_Eliece}, \cite{Poisel} when the jammer is not intelligent) is significantly lower than the performance of JB. 
The algorithms behave along similar lines in a non-coherent scenario where there is a random unknown phase offset between the jamming and the victim signals, as seen in Fig.~\ref{fig_non_coherent}. The jammer learned to use BPSK signaling at $\rho=0.051$ while the optimal jamming signal derived in \cite{GlobecomJamming} indicates that $\rho^*=0.06$ when $JNR=10\ dB$ and $SNR=20\ dB$. 

Now that we have established the performance of the proposed learning algorithm by comparing with previously known results, we now consider the performance of the learning algorithm in terms of the $PER$ which is a more relevant and practical metric to be considered in wireless environments. Further, it is also easy for the jammer to estimate $PER$ by observing the ACKs/NACKs exchanged between victim receiver and transmitter via the feedback channel \cite{Patent_PER}. Fig.~\ref{fig:PER_BPSK} shows the learning performance of various algorithms in terms of the average $PER$ inflicted by the jammer at the victim receiver. While the jammer learns to use BPSK as the optimal signaling scheme, the optimal $\rho$ value learned in this case is $0.23$ which is different from the value of $\rho$ learned in Fig.~\ref{fig1}. This is because $PER$ is used as the cost function in learning the jamming strategies. It is clear that both the AWGN jamming and $\epsilon$-greedy learning algorithm (that uses a sub-optimal value of $M$) achieve a $PER=0$ based on the $SER$ results in Fig.~\ref{fig1}. Even in this case, JB outperforms traditional jamming techniques that use AWGN or the $\epsilon$-greedy learning algorithm.

\begin{figure}
\begin{minipage}[t]{0.5\textwidth}
\centering
\includegraphics[width=2.5in]{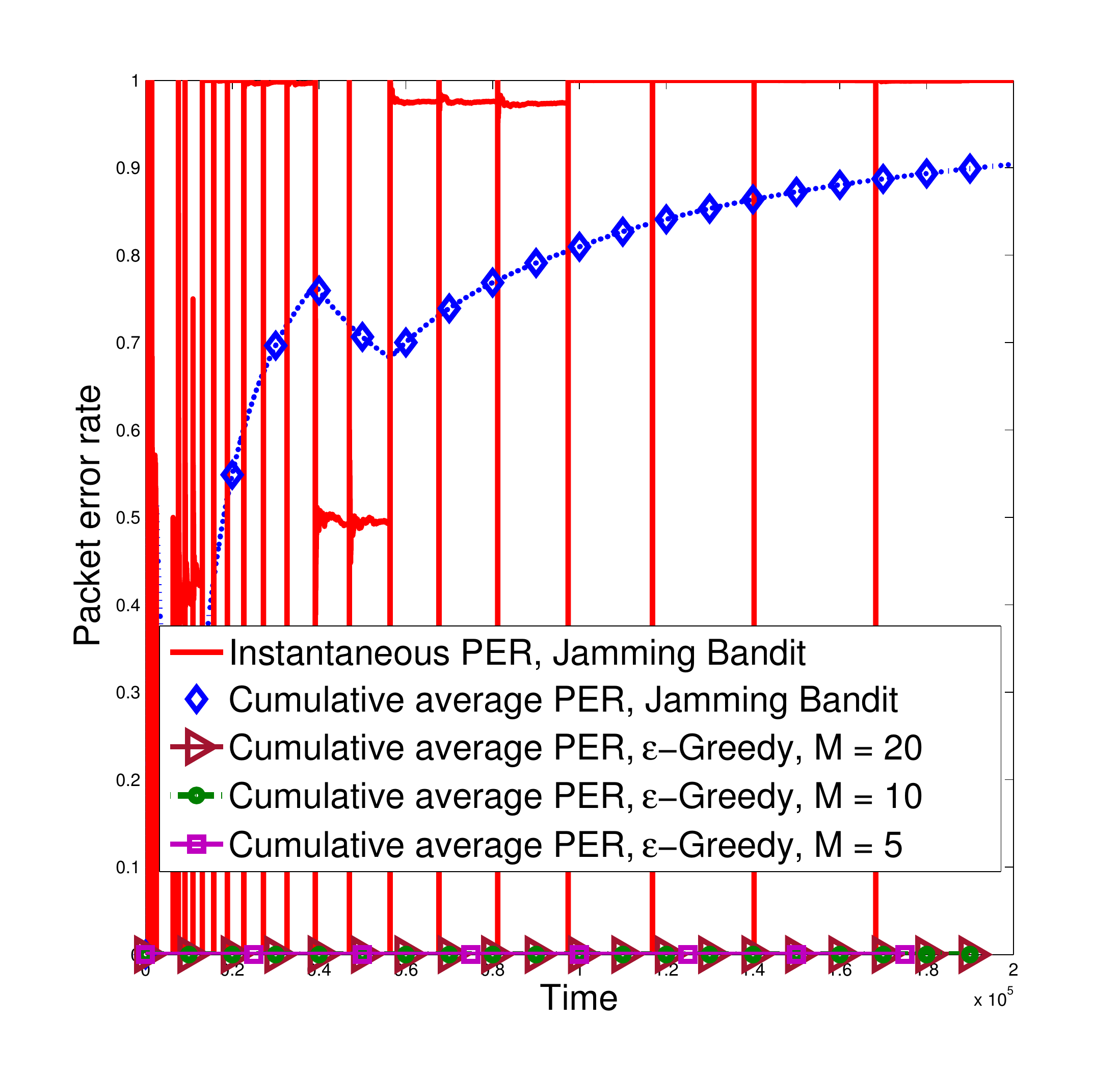}
\vspace{-10pt}
\caption{Average $PER$ inflicted by the jammer at the victim receiver, $\mathrm{SNR}=20$\ dB, victim uses BPSK and $\mathrm{JNR}=10$\ dB. The jammer learns to use BPSK signaling scheme with $\rho=0.23$.}
\label{fig:PER_BPSK}
\end{minipage}\hspace{8pt}
\begin{minipage}[t]{0.5\textwidth}
\centering
\includegraphics[width=2.5in]{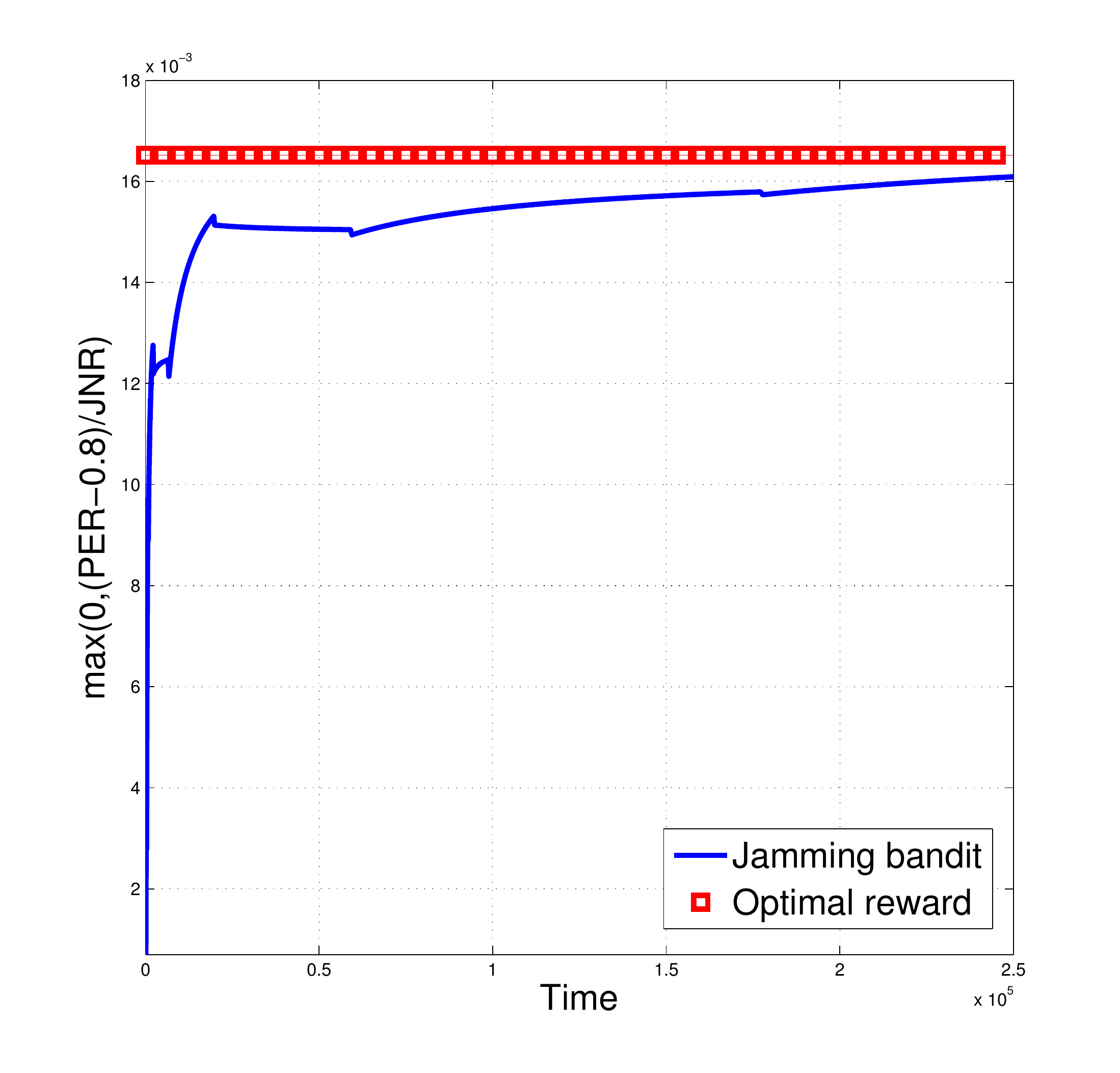}
\vspace{-10pt}
\caption{Average reward obtained by the jammer against a BPSK modulated victim, $SNR=20$\ dB. The optimal reward is obtained via grid search with discretization $M=100$.}
\label{fig:reward_PER_BPSK}
\end{minipage}
\vspace{-20pt}
\end{figure}

The cost function is taken as $\max(0,(PER(t)-0.8)/\mathrm{JNR}(t))$ (the cost function remains to be H\"{o}lder continuous and is bounded in $[0,1]$) to ensure that we choose only those strategies which achieve at least $80\%$ PER (remember, the jammer intends to maximize this cost/objective function). Fig.~\ref{fig:reward_PER_BPSK} compares the learning performance of JB with respect to the optimal strategy and Fig.~\ref{fig:confidence_level} shows the confidence levels as predicted by the one-step regret bound in Theorem~\ref{TheoremOneStepRegret} and that achieved by JB. The optimal reward is estimated by performing an extensive grid search $(M=100)$ over the entire strategy set. The steps in $\mathrm{log}\delta$ seen in Fig.~\ref{fig:confidence_level} are due to change in the discretization $M$ as shown in Algorithm~\ref{alg:1}. As mentioned before, the algorithm performs much better than predicted by the high confidence bound (evidenced by a lower value of $\delta$). 

\begin{figure}
\begin{minipage}[t]{0.5\textwidth}
\centering
\includegraphics[width=2.5in]{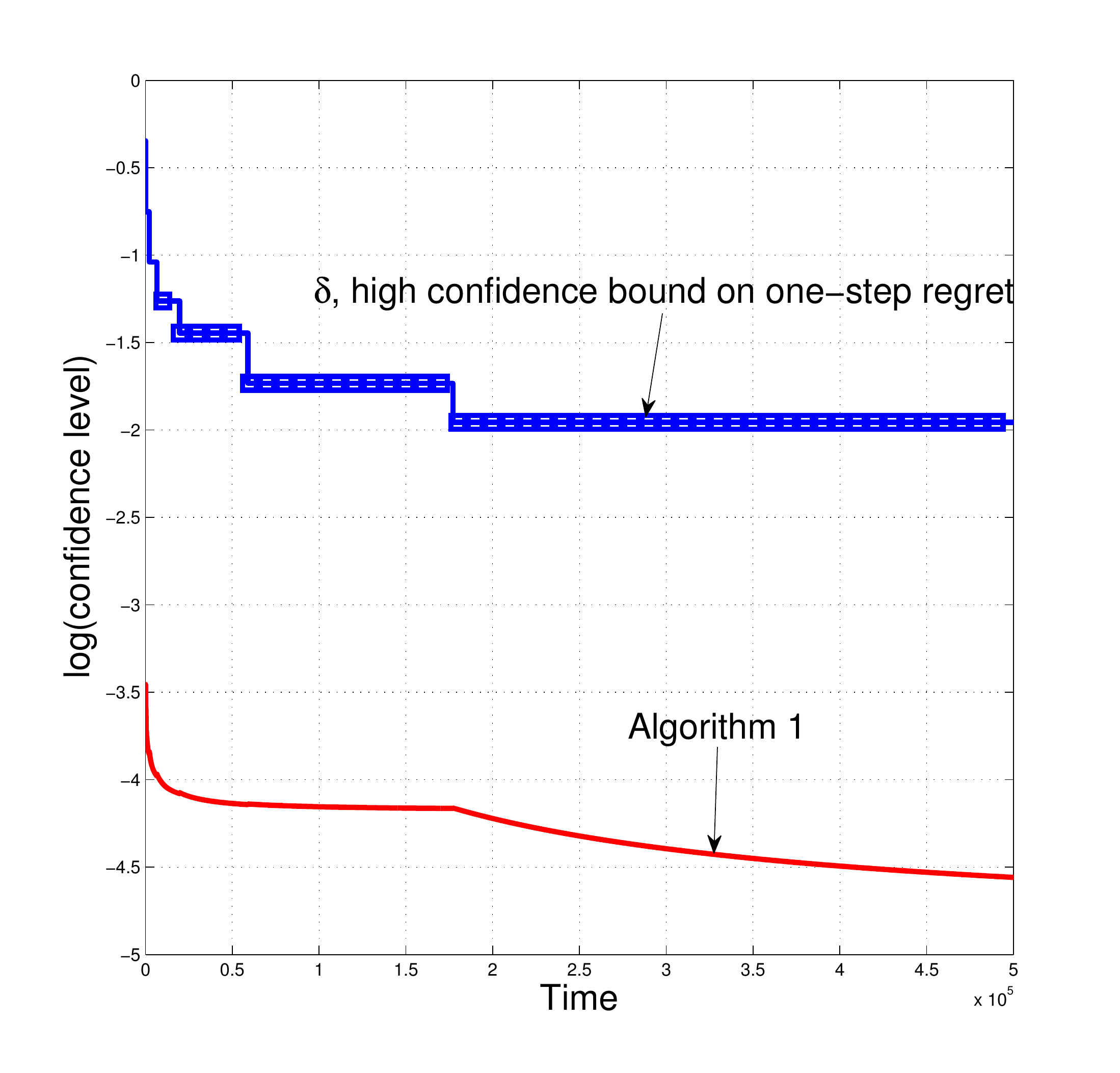}
\vspace{-10pt}
\caption{Confidence level (optimal reward-achieved reward) predicted by Theorem~\ref{TheoremOneStepRegret} and that achieved by JB.}
\label{fig:confidence_level}
\end{minipage}\hspace{5pt}
\begin{minipage}[t]{0.5\textwidth}
\centering
\includegraphics[width=2.5in]{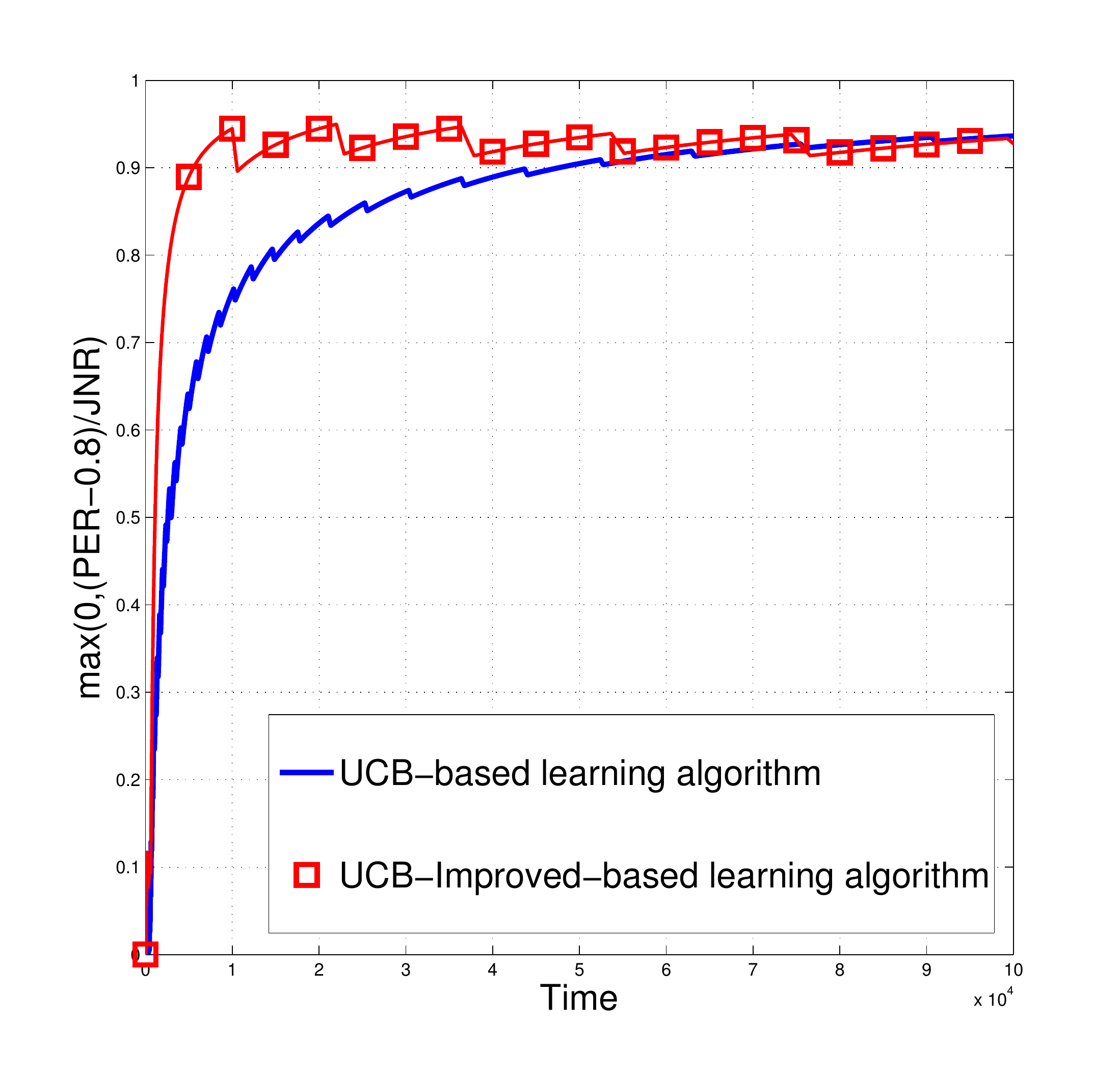}
\vspace{-10pt}
\caption{Learning the jamming strategies by using arm-elimination. The victim uses BPSK with $\mathrm{SNR}=20$dB. The jammer learned to use BPSK with $\mathrm{JNR}=15$\ dB and $\rho=0.22$. }
\label{figelim}
\end{minipage}
\vspace{-20pt}
\end{figure}

Fig.~\ref{figelim} shows the learning results obtained by using Algorithm~\ref{alg:2} i.e., JB uses the UCB-Improved algorithm in the inner loop instead of the UCB1 algorithm. It shows the learning performance of Algorithms~\ref{alg:1} and \ref{alg:2} in one inner loop iteration when $T=10^5$ (i.e., for one value of discretization $M$ evaluated as shown in Algorithm~\ref{alg:1}). It is seen that the Algorithm~\ref{alg:2} converges faster in comparison to the earlier approach as the algorithm eliminates sub-optimal arms and thereby only exploits the best jamming strategy. Even in this case the jammer learned to use BPSK signaling scheme against a BPSK-modulated victim signal. Further notice that the algorithm converges in about $10000$ time steps in this case as opposed to $>50000$ time steps using JB. Recall that in the simulations we assume that one packet is sent every time instant and hence in order to obtain reliable estimates of the performance of each jamming strategy, the jammer requires about $10000$ time instants. 
\vspace{-10pt}
\subsection{Jamming Performance Against an Adaptive Victim}
We first assume that the victim employs a uniform distribution over its strategy set i.e., it chooses uniformly at random (at every time instant) a power level in the range $[\mathrm{SNR}_{\min},\mathrm{SNR}_{\max}]$ and the modulation scheme from the set $\{$BPSK,QPSK$\}$. The performance of JB when the victim employs such a stochastic strategy is shown in Fig.~\ref{fig3}. Again, the superior performance of the bandit-based learning algorithm when compared to the traditionally used AWGN jamming and naive learning algorithms such as $\epsilon$-Greedy is proved from these results.\footnote{It is worth noticing that model-free learning algorithms such as Q-Learning and SARSA \cite{CNSQLearning} cannot be employed in the scenarios considered in this paper because it is assumed that the jammer cannot observe any of the environment parameters such as the victim's modulation scheme and power levels. However, it is expected that the performance of the learning algorithms can be improved when such additional information is available, which is typically the case in optimization-based algorithms.} 
 
\begin{figure}
\centering
\vspace{-20pt}
\includegraphics[width=2.5in]{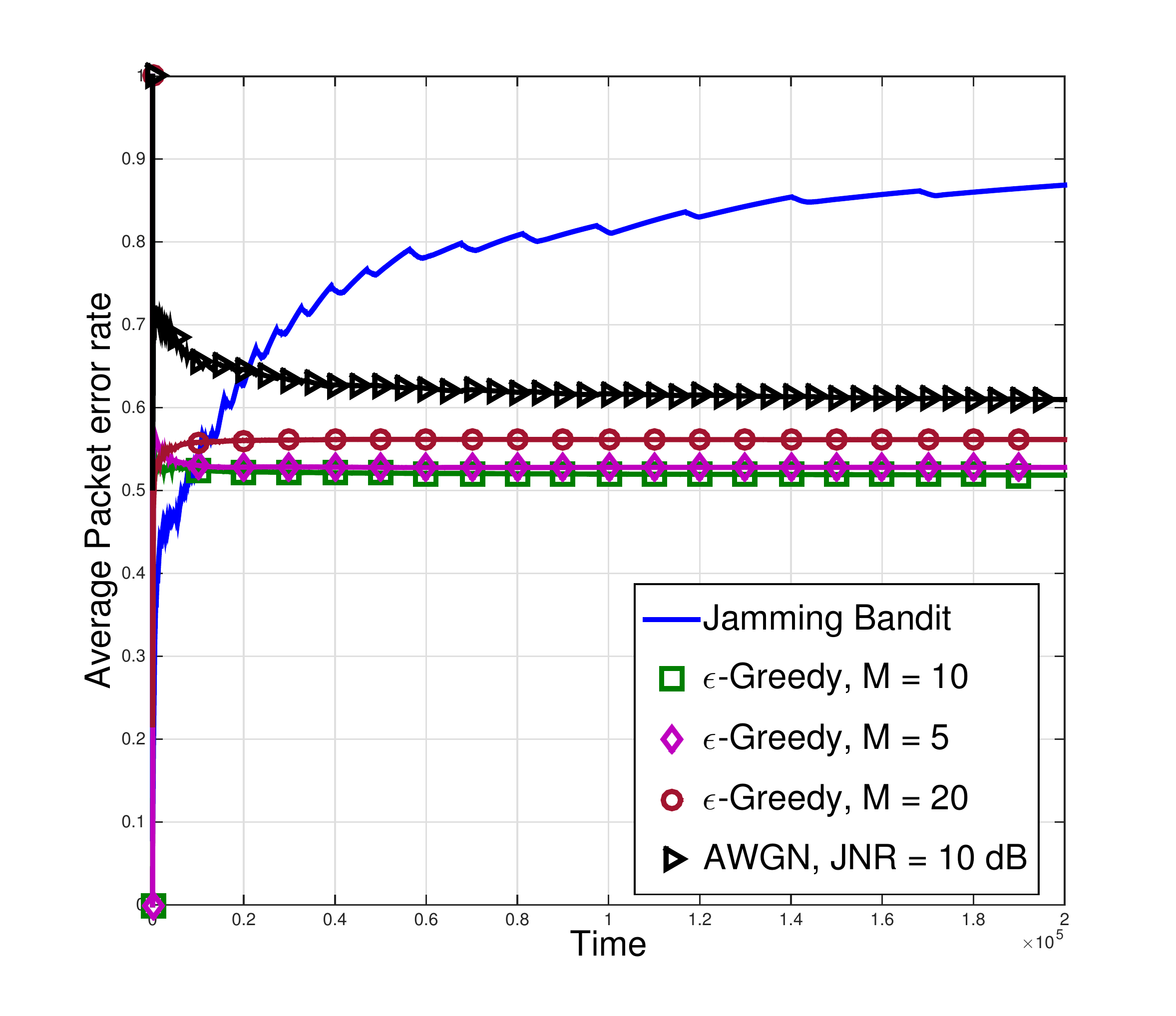}
\vspace{-20pt}
\caption{Learning jammers' strategy against a stochastic user. The victim transmitter-receiver pair use a uniformly random signaling scheme that belongs to the set $\{$BPSK,QPSK$\}$ and random power level in the range $[0,20]$\ dB.}
\label{fig3}
\end{figure}

\begin{figure}
\begin{minipage}[t]{0.5\textwidth}
\vspace{-15pt}
\centering
\includegraphics[width=2.5in]{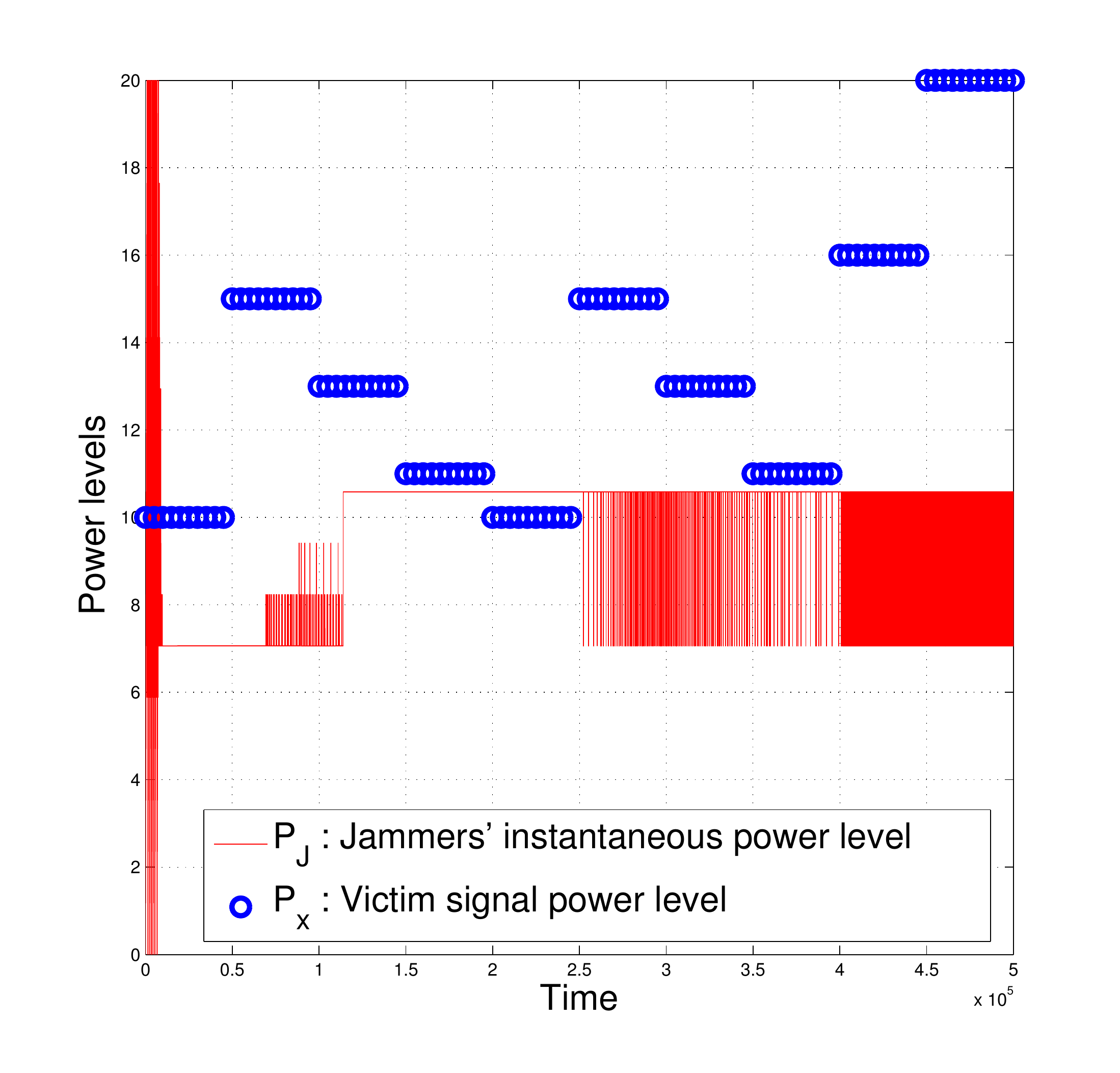}
\vspace{-22pt}
\caption{Learning against a victim with time-varying strategies. The figure shows the power levels adaptation by the jammer and that used by the victim.}
\label{fig:No_Drifting_Performance_Stochastic_User}
\end{minipage}\hspace{5pt}
\begin{minipage}[t]{0.5\textwidth}
\centering
\vspace{-15pt}
\includegraphics[width=2.5in]{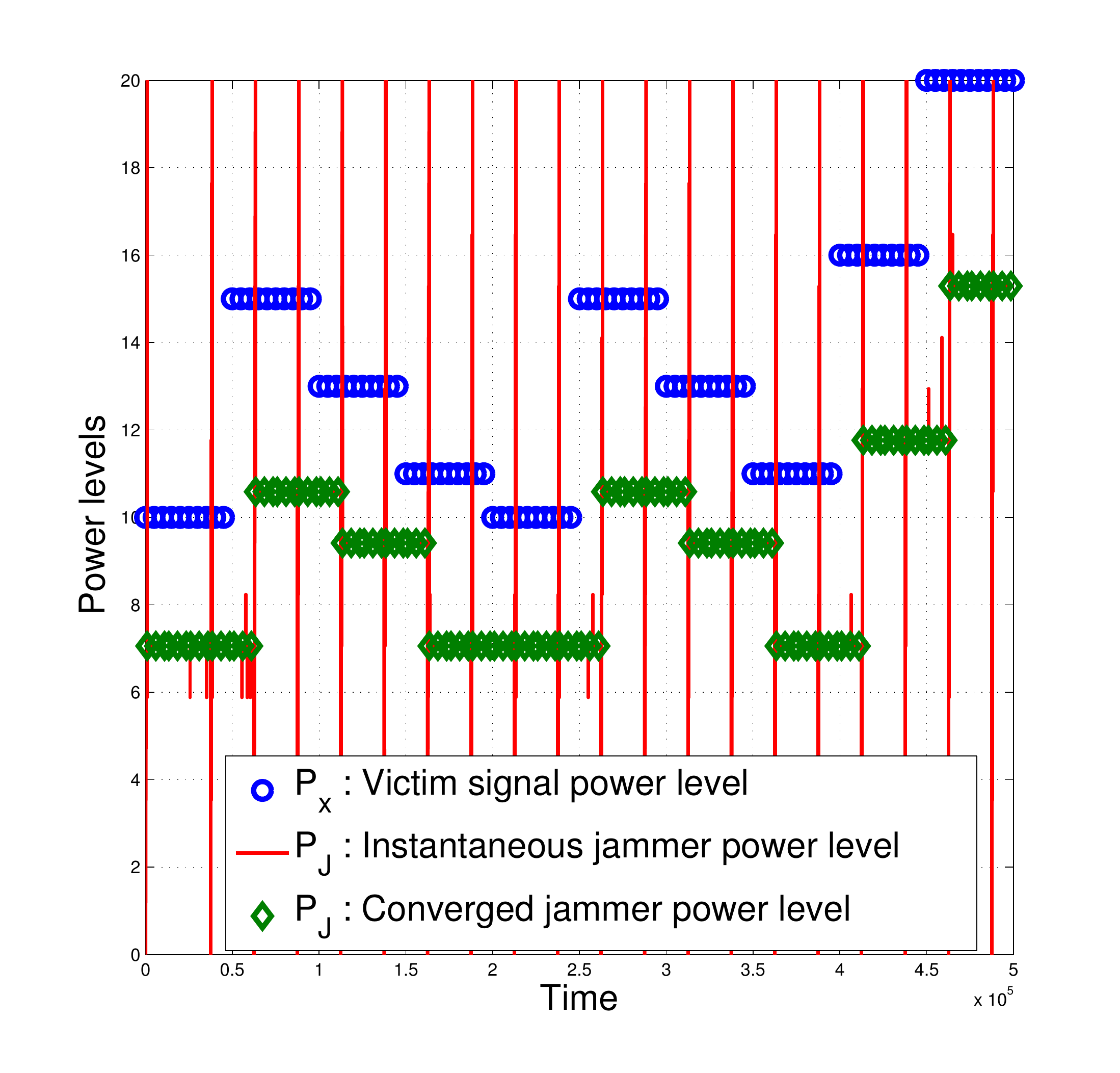}
\vspace{-20pt}
\caption{Learning against a victim with time-varying strategies. The figure shows the power levels adaptation by the jammer using a drifting algorithm and that used by the victim.}
\label{fig:Drifting_Performance_Stochastic_User}
\end{minipage}
\vspace{-22pt}
\end{figure}

When the victim changes its strategy rapidly, JB cannot track the changes perfectly as seen in Fig.~\ref{fig:No_Drifting_Performance_Stochastic_User}  because it learns over all past information, and prior information may not convey knowledge about the current strategy used by the victim which can be completely different from the prior strategy. In such cases, it is important to learn only from recent past history, which can be achieved by using JB on a recent window of past history (for instance, a sliding window-based algorithm to track changes in the environment) \cite{CemDrifting}. Specifically, we use the concept of drifting \cite{CemDrifting}
to adapt to the victim's strategy.
In this algorithm, each round $i$ (which is of $T$ time steps, where $T=2^i$) is divided it into several frames each of $W$ time instants. Within each frame, the first $W/2$ time steps, are termed as the passive slot and the second $W/2$ time instants are termed as the active slot. In the first frame, both the slots will be taken to be active slots. Each passive slot overlaps with the active slot of the previous frame. If time $t$ belongs to active slot of frame $w$, then actions are taken as per the UCB1 indices evaluated in this particular frame $w$. However, if it belongs to the passive slot of frame $w$, which is taken to overlap with the active slot of frame $w-1$, then it takes actions as per the indices of the frame $w-1$, but updates the UCB1 indices so that it can be used in frame $w$. Specifically, at the start of every frame $w$, the counters and mean reward estimates are all reset to zero and when actions are taken in the passive slot of frame $w$, these counters and reward estimated are updated so as to be used in the active slot. Thus when the algorithm enters the active slot of frame $w$, it already has some observations using which it can exploit without wasting time in the exploration phase. Such splitting of the time horizon will enable the jammer to quickly adapt to the victim's varying strategies. Please see \cite{CemDrifting} for more details on the drifting algorithm. Specifically, we consider the drifting algorithm with a window length $W=25000$.
 
Fig.~\ref{fig:Drifting_Performance_Stochastic_User} shows the jammers' power level adaption when the victim is randomly varying its power levels across time and the jammer employs the drifting algorithm in conjunction with JB. The dips seen at regular intervals in Fig.~\ref{fig:Drifting_Performance_Stochastic_User} are due to the proposed sliding window-based algorithm where the user resets the algorithm at regular intervals to adapt to the changing wireless environment. The $PER$ achieved by this algorithm is similar to the results shown in Figs.~\ref{fig:PER_BPSK}, \ref{fig:confidence_level} in comparison to other jamming techniques. While Fig.~\ref{fig:Drifting_Performance_Stochastic_User} considered the case when the victim changes its power levels randomly, the jammer can also easily track the victim when it employs commonly used adaption strategies such as increasing the power levels when $PER$ increases and vice versa. These results successfully illustrate the adaptive capabilities of the proposed learning algorithms that can overcome the difficulties faced by JB as shown in Fig.~\ref{fig:No_Drifting_Performance_Stochastic_User}. 

\vspace{-11pt}
\subsection{Multiple Victims}\label{sec:MultipleUsers}
In this subsection, we consider a case when the jammer uses an omnidirectional antenna and intends to jam two victims in a network. Interesting scenarios arise in this scenario because the jammer has to optimize its jamming strategy based on the $PER$ of both the victims. For example, when both the victims use BPSK, the jammer will learn to use BPSK signaling scheme but the power level at which it should jam depends on the relative power levels of both the victims. Several factors such as path loss, shadowing etc. akin to practical wireless systems can be introduced into this problem, but we are mainly interested in understanding the learning performance of the jammer. Hence we ignore these physical layer parameters and assume that both the victims are affected by the jamming signal with the same $\mathrm{JNR}$.
The jammer considers the mean packet error rate seen at both these victims as feedback with target mean $PER=0.8$, in order to learn the performance of its actions. 
\begin{figure}
\begin{minipage}[t]{0.5\textwidth}
\centering
\includegraphics[width=2.5in]{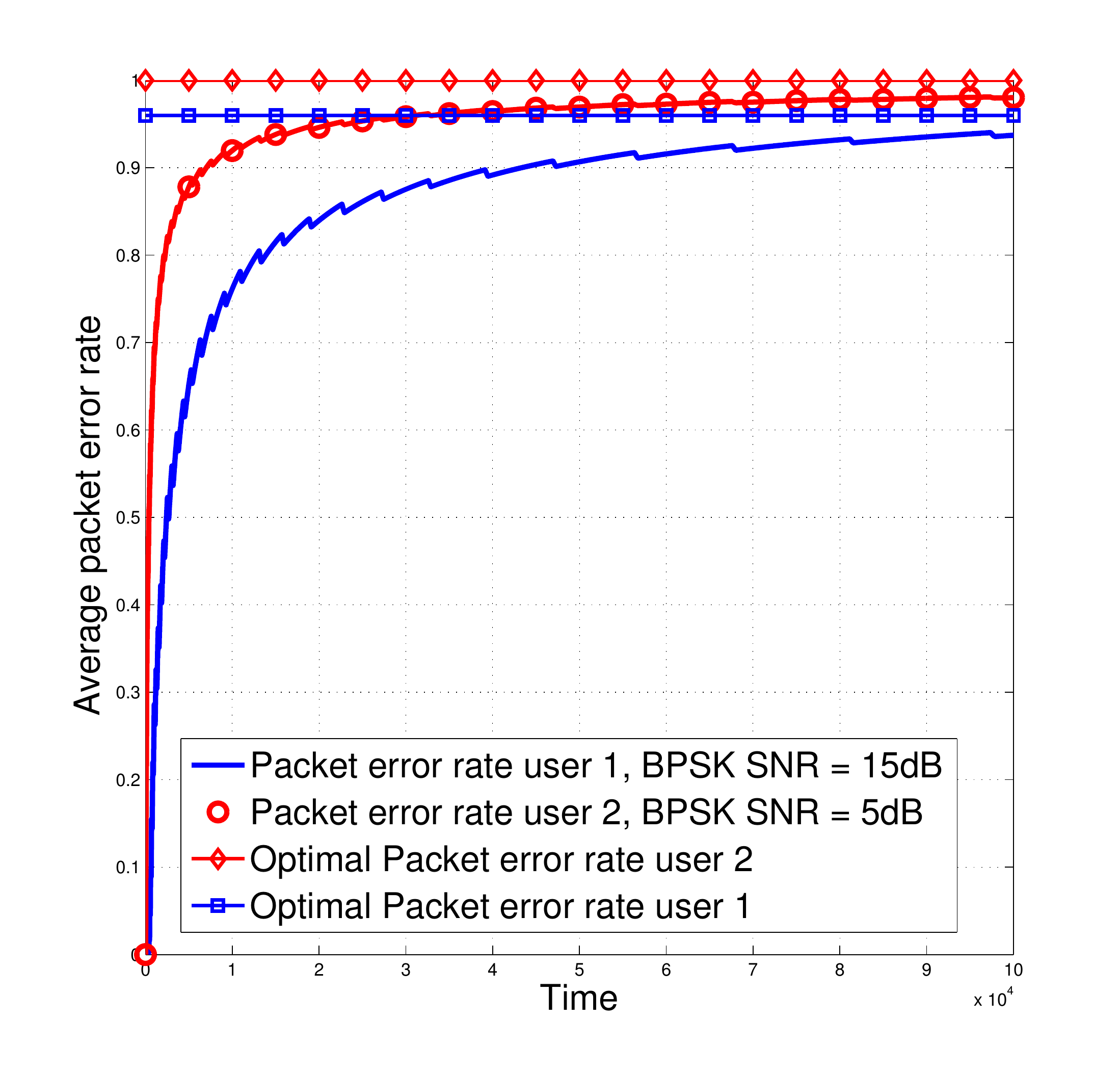}
\vspace{-20pt}
\caption{PER achieved by the jammer against 2 users, user 1 uses BPSK at $15$dB and user 2 sends BPSK at $5$dB. The jammer learns to use BPSK signal with power $13$dB and $\rho=0.46$.}
\label{MU_BPSK}
\end{minipage}\hspace{5pt}
\begin{minipage}[t]{0.5\textwidth}
\centering
\includegraphics[width=2.5in]{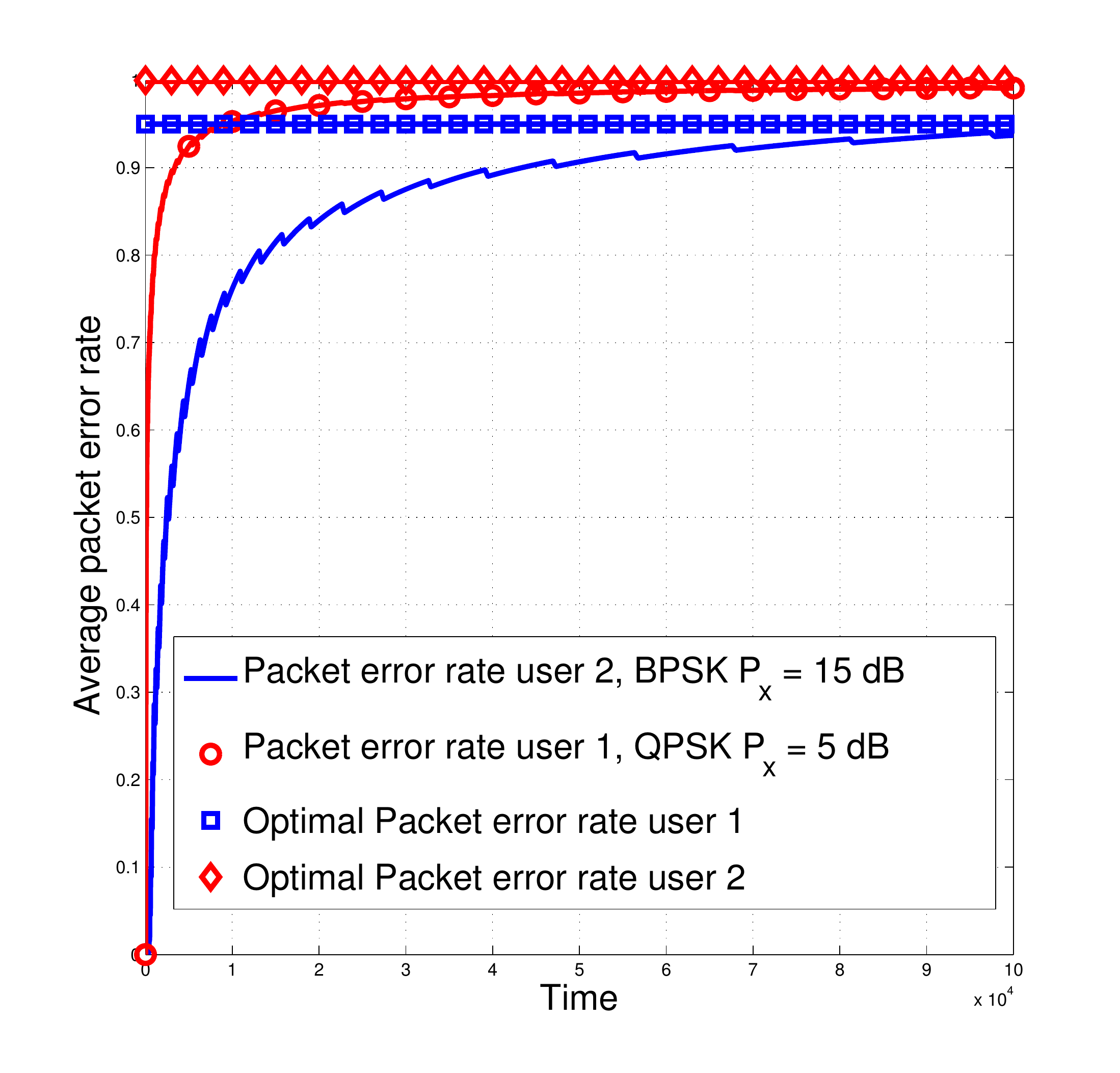}
\vspace{-20pt}
\caption{PER achieved by the jammer against 2 users, user 1 sends QPSK at $5$dB and user 2 sends BPSK at $15$dB. The jammer learns to use BPSK signal with power $11.25$dB and $\rho=0.25$.}
\label{MU_QPSK_BPSK}
\end{minipage}
\vspace{-20pt}
\end{figure}

Fig.~\ref{MU_BPSK} shows the learning performance of the jammer against $2$ users that employ BPSK signaling at different power levels. It is seen that the jammer learns to use BPSK signaling as well (since BPSK is optimal to be used against BPSK signaling as discussed in \cite{GlobecomJamming}). Similar learning results were achieved when both the users employ QPSK signaling. 
Figs.~\ref{MU_QPSK_BPSK} shows the learning performance when one user uses QPSK and and the other user uses BPSK. It was observed that when the victim with BPSK has higher power than QPSK victim, the jammer learns to use the BPSK jamming signal and vice versa. This again agrees with previous results which show that BPSK (QPSK) is better to jam a BPSK (QPSK) signal. Also, the learning algorithm performs comparably well to the optimal strategy obtained by performing an extensive grid search over the complete set of strategies. Fig.~\ref{MU_Drifting} shows the performance of the JB algorithm against the two users that are randomly changing their power levels to overcome interference  (this captures a much more difficult scenario as compared to standard adaptive mechanisms, such as power control schemes, in which the victim increases its power level until it reaches a maximum so as to overcome interference). Although each victim has a different adaption cycle (specifically, victim $1$ changes its power levels based on the performance history over the past $50000$ time instants and victim $2$ adapts its power levels over a window of size $30000$ time instants), the jammer is capable of tracking these changes in a satisfactory manner. 

%

\begin{figure}
\centering
\includegraphics[width=2.5in]{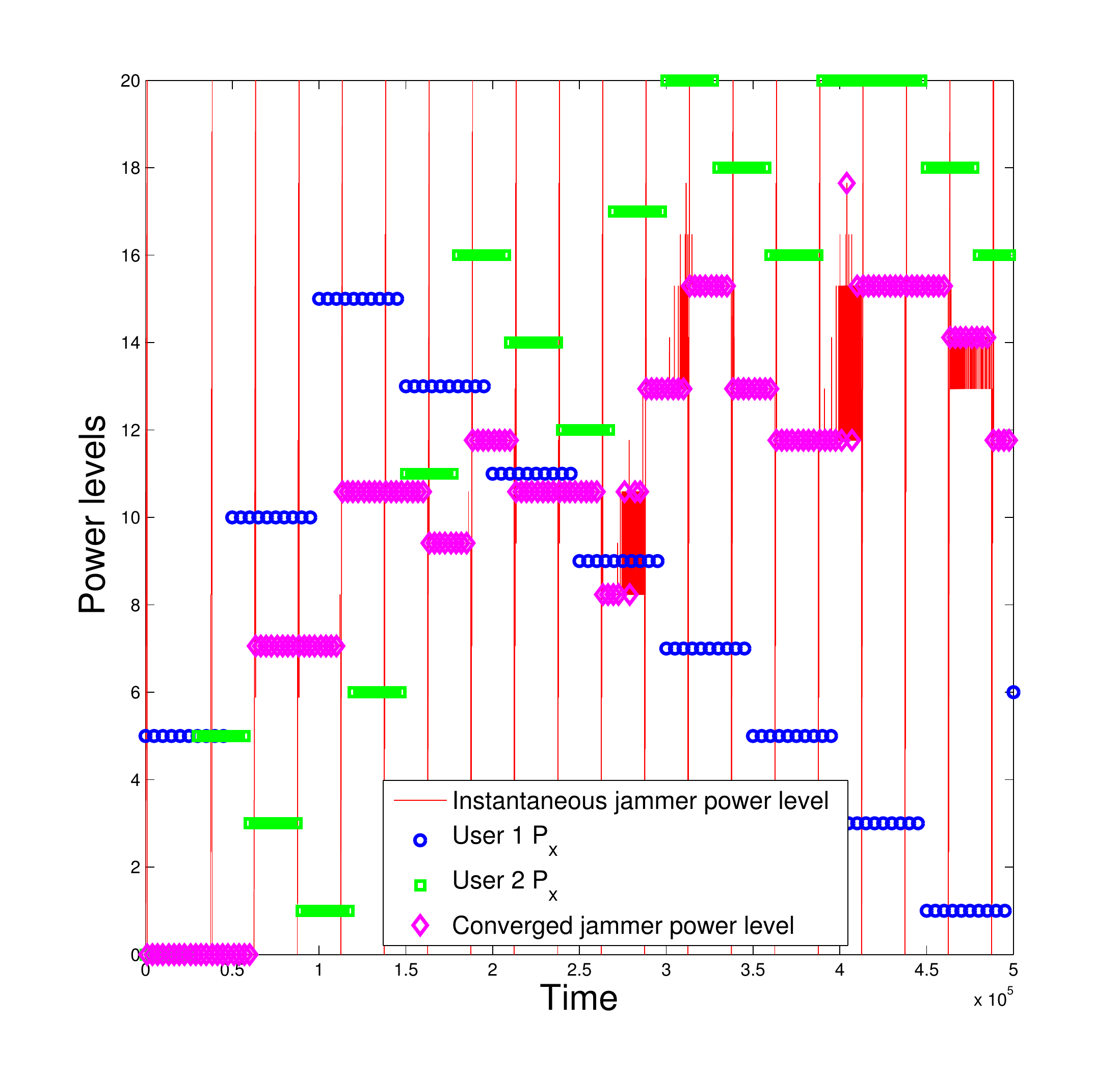}
\vspace{-20pt}
\caption{PER achieved by the jammer against 2 stochastic users in the network. Both the users employ BPSK signaling scheme. The jammer learns to use the BPSK signaling scheme to achieve power efficient jamming strategies and also tracks the changes in the users' strategies. }
\label{MU_Drifting}
\vspace{-20pt}
\end{figure}

By using a weighted $PER$ metric rather than a mean $PER$ metric, the jammer can prioritize jamming one set of transmit-receive pairs against the others. Several other MU cases can easily be considered by using this framework. For example, by allowing the jammer to choose the direction of jamming as another action, the jammer can prioritize jamming only the transmit-receive pairs in a given direction rather than spread all its power uniformly across all directions. However, such improved jamming techniques will only come at the expense of more knowledge about the location of the users, users' behavior etc. Nevertheless, it is worth appreciating the applicability of the proposed algorithms to a wide variety of electronic warfare-type scenarios. 
\vspace{-11pt}
\section{Conclusion}\label{Conclusions}
In this paper, we proved that a cognitive jammer can learn the optimal physical layer jamming strategy in an electronic warfare-type scenario without having any \emph{a priori} knowledge about the system dynamics. Novel learning algorithms based on the multi-armed bandit framework were developed to optimally jam victim transmitter-receiver pairs. The learning algorithms are capable of learning the optimal jamming strategies in both coherent and non-coherent scenarios where the jamming signal and the victim signal are either phase synchronous or asynchronous with each other. Also, the rate of learning is faster in comparison to commonly used reinforcement learning algorithms. These algorithms are capable of tracking the different strategies used by multiple adaptive transmitter-receiver pairs. Moreover, they come with strong theoretical guarantees on the performance including confidence bounds which are used to estimate the probability of successful jamming at any time instant. 

{\linespread{1}

}
\vspace{-11pt}
\section*{Appendix A \\ Proof of Theorem~\ref{Lipschitz_Theorem}}\label{AppendixLipschitz}
For the system model in Section~\ref{sec:SystemModel}, the average probability of error at the victim receiver that uses a maximum likelihood (ML) detector (since it is assumed that the victim transmit-receive pair is not aware of the presence of the jammer) is given by
{\small \begin{align}\label{BER_initial}
p_e\left(j,\mathrm{SNR},\mathrm{JNR}\right)&=1-\int_x\int_{\Omega}f_N\left(y-\sqrt{\mathrm{SNR}}x-\sqrt{\mathrm{JNR}}j\right)f_X(x)dydx,
\end{align}}
where $\Omega$ indicates the ML decision region for $x$. For instance, when the signal levels are $\pm A$, $\Omega=\mathsf{real}\left(y\right)<0$ when $x=-A$ and $\Omega=\mathsf{real}\left(y\right)>0$ when $x=+A$. In the above equation, the received signal normalized by the noise power $\sigma^2$ is considered. Further, $f_X$ indicates the distribution of the signal $x$ (described by the modulation scheme used by the victim) and $f_N$ indicates the additive white Gaussian noise distribution. For a pulsed jamming signal with pulsing ratio $\rho$, the $SER$ is given by $\rho p_e(j,\mathrm{SNR},\frac{\mathrm{JNR}}{\rho})+(1-\rho)p_e(j,\mathrm{SNR},0)$. We first establish the H\"{o}lder continuity of $p_e\left(j,\mathrm{SNR},\mathrm{JNR}\right)$ which can then be used to prove the H\"{o}lder continuity of $p_e$ for pulsed jamming scenarios. 

In order to prove that $SER$ i.e., $p_e$ is uniformly locally Lipschitz, we show that $|p_e(j,\mathrm{SNR},\mathrm{JNR}_1)-p_e(j,\mathrm{SNR},\mathrm{JNR}_2)|\leq L|\mathrm{JNR}_1-\mathrm{JNR}_2|^\alpha$ for some $L>0$ and $\alpha>0$. Using \eqref{BER_initial} we have,
{\small \begin{align}\label{intermediate_Lipschitz}
&|p_e(j,\mathrm{SNR},\mathrm{JNR}_1)-p_e(j,\mathrm{SNR},\mathrm{JNR}_2)|\nonumber \\
&\hspace{-10pt}=|\int_x\int_{\Omega}\left[f_N\left(y-\sqrt{\mathrm{SNR}}x-\sqrt{\mathrm{JNR}_2}j\right)-f_N\left(y-\sqrt{\mathrm{SNR}}x-\sqrt{\mathrm{JNR}_1}j\right)\right]f_X(x)dydx|\nonumber \\
&\hspace{-10pt}\leq\int_x\int_{\Omega}\left[|f_N\left(y-\sqrt{\mathrm{SNR}}x-\sqrt{\mathrm{JNR}_2}j\right)-f_N\left(y-\sqrt{\mathrm{SNR}}x-\sqrt{\mathrm{JNR}_1}j\right)|\right]f_X(x)dydx.
\end{align}}
Thus it is sufficient to show that $|f_N\left(y-\sqrt{\mathrm{SNR}}x-\sqrt{\mathrm{JNR}_2}j\right)-f_N\left(y-\sqrt{\mathrm{SNR}}x-\sqrt{\mathrm{JNR}_1}j\right)|\leq L'|\mathrm{JNR}_1-\mathrm{JNR}_2|^{\alpha'}$ for some
$L'>0$ and $\alpha'>0$ which follows from the definition of $f_N$ as it is the probability density function (pdf) of the noise signal $n$. We briefly show it below for completeness. Since we already normalized the signal by $\sigma^2$, the pdf of $n$ is now given by the zero mean unit variance Gaussian distribution. 
{\small \begin{align}\label{Lipschitz_intermediate}
|f_N\left(y-\sqrt{\mathrm{SNR}}x-\sqrt{\mathrm{JNR}_2}j\right)-f_N\left(y-\sqrt{\mathrm{SNR}}s-\sqrt{\mathrm{JNR}_1}j\right)|&\nonumber \\
&\hspace{-280pt}=|\frac{1}{\sqrt{2\pi}}\left[\exp(-(y-\sqrt{\mathrm{SNR}}x-\sqrt{\mathrm{JNR}_2}j)^2)-\exp(-(y-\sqrt{\mathrm{SNR}}x-\sqrt{\mathrm{JNR}_1}j)^2)\right]|\nonumber \\
&\hspace{-280pt}\approx \frac{1}{\sqrt{2\pi}}\left[|(\sqrt{JNR}_1-\sqrt{JNR}_2)j|\right],
\end{align}}
where the last approximation is obtained by ignoring the higher order terms since we only consider the cases where $|\mathrm{JNR}_1-\mathrm{JNR}_2|\leq \delta$ i.e., local H\"{o}lder continuity. Then, for a given jamming signal $j$, \eqref{Lipschitz_intermediate} can be bounded as 
\vspace{-5pt}
{\small \begin{align}\label{argument_Lipschitz}
\frac{1}{\sqrt{2\pi}}\left[|(\sqrt{JNR}_1-\sqrt{JNR}_2)j|\right]&\leq\sqrt{\frac{\mathrm{JNR}_2}{2\pi}}|\left(\sqrt{1+\frac{\delta}{\mathrm{JNR}_2}}-1\right)|\approx \sqrt{\frac{\mathrm{JNR}_2}{2\pi}}|\left(1+\frac{\delta}{2\mathrm{JNR}_2}-1\right)| \nonumber \\
&\leq \sqrt{\frac{1}{2\pi \mathrm{JNR}_{\min}}}\delta =\sqrt{\frac{1}{2\pi \mathrm{JNR}_{\min}}}(\mathrm{JNR}_1-\mathrm{JNR}_2),
\end{align}}
which proves that argument inside the integral in \eqref{intermediate_Lipschitz} is uniformly locally Lipschitz with $L'=\sqrt{\frac{1}{2\pi \mathrm{JNR}_{\min}}}$ and $\alpha'=1$. In the above proof we used the fact that $|j|\leq 1$ for standard signaling schemes that are employed by the jammer (for the AWGN jamming signal, the $SER$ is obtained by using a Gaussian distribution with variance $1+\mathrm{JNR}$ i.e., a slightly different approach when compared to \eqref{BER_initial} is taken and by following the above sequence of arguments, H\"{o}lder continuity can be proved even in this case). Using \eqref{argument_Lipschitz}, the overall $SER$ i.e., $p_e(j,\mathrm{SNR},\mathrm{JNR})$ is also uniformly locally H\"{o}lder continuous. By following the same steps, the H\"{o}lder continuity for the pulsed jamming cases can also be proved. An example for the H\"{o}lder continuity in the pulsed jamming case is shown in Section~\ref{sec:FixedUser}.

\vspace{-11pt}
\section*{Appendix B \\Proof of Theorem~\ref{theorem1}}\label{AppendixA}
Since the set of signaling schemes is discrete, we first obtain the regret bound for a particular signaling scheme $\mathcal{J}$. It is easy to see that the overall regret bound 
is a scaled version (by $N_{mod}$) of the regret achievable for a single signaling scheme. Since the time horizon of the inner loop of  Algorithm~\ref{alg:1} is $T$, we first show that the regret incurred by the inner loop is $\mathcal{O}(\sqrt{M^2T \mathrm{log}(T)})$. Since the overall time horizon is generally unknown, the algorithm is run for several rounds of time steps on the order of $2^i$ as shown in Algorithm~\ref{alg:1} and the regret bounds for the overall algorithm can be achieved by using the doubling trick\cite{Audibert}. 

The upper bound on the overall regret incurred by Algorithm~\ref{alg:1} can be obtained by upper bounding $\sum_{t=1}^T\left(\bar{C}(\mathcal{J},\mathbf{s}^*)-\bar{C}(\mathcal{J},\mathbf{s}_t)\right)$, 
where $\bar{C}$ indicates the average cost function and $\mathbf{s}^*$ is the best strategy for a given signaling scheme $\mathcal{J}$ and $\mathbf{s}_t$ is the actual strategy chosen at time $t$. For ease of presentation, $\mathcal{J}$ is ignored in the rest of the proof. We obtain the regret bound in two steps by rewriting it as 
\begin{align}\label{cumulative_average_regret}
\sum_{t=1}^T\left(\bar{C}(\mathbf{s}^*)-\bar{C}(\mathbf{s}_t)\right)=\sum_{t=1}^T\left(\bar{C}(\mathbf{s}^*)-\bar{C}(\mathbf{s}')\right)+\sum_{t=1}^T\left(\bar{C}(\mathbf{s}')-\bar{C}(\mathbf{s}_t)\right),
\end{align}
where $\mathbf{s}'\in \{1/M,2/M,\ldots,1\}\times \mathrm{JNR}_{\min}+(\mathrm{JNR}_{\max}-\mathrm{JNR}_{\min})*\{1/M,2/M,\ldots,1\}$ is the strategy nearest (in terms of the Euclidean distance) to $\mathbf{s}^*$. Then we have $||\mathbf{s}'-\mathbf{s}^*||$=$\sqrt{(\mathrm{JNR}'-\mathrm{JNR}^*)^2+(\rho'-\rho^*)^2}$ $\leq\sqrt{\frac{2}{M^2}}$ based on the discretization of the continuous arms set in Algorithm~\ref{alg:1}.

For the first term in the above equation, by using the H\"{o}lder continuity properties of the average cost function $\bar{C}$, we have \vspace{-10pt}
\begin{align}\label{regret_term_1}
\mathbf{E}\Big(\sum_{t=1}^TC_t(\mathbf{s}^*)-C_t(\mathbf{s}')\Big)&=\sum_{t=1}^T\left(\bar{C}(\mathbf{s}^*)-\bar{C} (\mathbf{s}')\right)\leq TL\Big(\frac{2}{M^2}\Big)^{\alpha/2}.
\end{align}
We now bound the second term $\mathbf{E}\Big(\sum_{t=1}^TC_t(\mathbf{s}')-C_t(\mathbf{s}_t)\Big)=\sum_{t=1}^T\left(\bar{C}(\mathbf{s}')-\bar{C}(\mathbf{s}_t)\right)$. Due to the discretization technique used in Algorithm~\ref{alg:1}, this problem is equivalent to a standard MAB problem with $M^2$ arms \cite{PeterAuer}. In order to bound \eqref{regret_term_1}, we define two sets of arms: near-optimal arms and sub-optimal arms. We set $\Delta=\sqrt{M^2\mathrm{log}(T)/T}$ and say that an arm is sub-optimal in this case, if its regret incurred is greater than $\Delta$ and near-optimal when its regret is less than $\Delta$. Thus, for a near-optimal arm, even when that arm is selected at all time steps, the contribution to regret will be at most $T\Delta$. In contrast for a sub-optimal arm, the contribution to the regret when it is selected can be large. Since we use the UCB1 algorithm, it can be shown that the sub-optimal arms will be chosen only $\mathcal{O}(\mathrm{log}(T)/\Delta(\mathbf{s})^2)$ times ($\Delta(\mathbf{s})$ is the regret of the strategy $\mathbf{s}$) \cite{PeterAuer}, before they are identified as sub-optimal. Thus the regret for these sub-optimal arms is on the order of $\mathcal{O}(\mathrm{log}(T)/\Delta)$ since $\Delta(\mathbf{s})>\Delta$.
From these arguments the second term in \eqref{cumulative_average_regret} can be upper bounded as
\begin{align} \label{regret_term_2}
\mathbf{E}\Big(\sum_{t=1}^TC_t(\mathbf{s}_t)-C_t(\mathbf{s}')\Big)\leq \mathcal{O}(\sqrt{M^2T\mathrm{log}(T)}).
\end{align}

Using \eqref{regret_term_1} and \eqref{regret_term_2}, and setting $M=\ceil{(\sqrt{\frac{T}{\mathrm{log} (T)}}L2^{\alpha/2})^{\frac{1}{1+\alpha}}}$ (this is obtained by matching the regret bounds shown in \eqref{regret_term_1} and \eqref{regret_term_2}), the regret for any given signaling scheme is given by $\mathcal{O}(\sqrt{M^2 T\mathrm{log}(T)})$. By noting the fact that the jammer can choose from $N_{mod}$ possible signaling schemes and using the value of $M$, the doubling trick, and summing the regret over all inner loop iterations of Algorithm~\ref{alg:1}, the regret over the entire time horizon $n$ can be expressed as $\mathcal{O}(N_{mod}{n^{\frac{\alpha+2}{2(\alpha+1)}} (\mathrm{log} n)^{\frac{\alpha}{2(\alpha+1)}}})$. 

\section*{Appendix C \\ High Confidence Bounds}
Here, we present high confidence bounds on the one-step and cumulative regret. 
\subsection{Proof of Theorem~\ref{TheoremOneStepRegret}}
We bound the one-step regret as follows,
\begin{align}\label{theorem_one_step_main_eq}
P(\bar{C}(\mathcal{J}^*,\mathbf{s}^*)-\bar{C}(\mathcal{J}_t,\mathbf{s}_t)>\delta)&=P(\bar{C}(\mathcal{J}^*,\mathbf{s}^*)-\bar{C}(\mathcal{J}_t,\mathbf{s}^*)+\bar{C}(\mathcal{J}_t,\mathbf{s}^*)-\bar{C}(\mathcal{J}_t,\mathbf{s}_t)>\delta) \nonumber \\
&\hspace{-120pt}=P(\bar{C}(\mathcal{J}^*,\mathbf{s}^*)-\bar{C}(\mathcal{J}_t,\mathbf{s}^*)+\bar{C}(\mathcal{J}_t,\mathbf{s}^*)-\bar{C}(\mathcal{J}_t,\mathbf{s}')+\bar{C}(\mathcal{J}_t,\mathbf{s}')-\bar{C}(\mathcal{J}_t,\mathbf{s}_t)>\delta) \nonumber \\
&\hspace{-120pt}\leq P(\bar{C}(\mathcal{J}^*,\mathbf{s}^*)-\bar{C}(\mathcal{J}_t,\mathbf{s}^*)\geq \frac{\delta}{4})+P(\bar{C}(\mathcal{J}_t,\mathbf{s}^*)-\bar{C}(\mathcal{J}_t,\mathbf{s}')\geq \frac{\delta}{2})\nonumber \\
&\hspace{50pt}+P(\bar{C}(\mathcal{J}_t,\mathbf{s}')-\bar{C}(\mathcal{J}_t,\mathbf{s}_t)\geq \frac{\delta}{4})
\end{align}

For the second term in \eqref{theorem_one_step_main_eq}, we have
\begin{align}
P(\bar{C}(\mathcal{J}_t,\mathbf{s}^*)-\bar{C}(\mathcal{J}_t,\mathbf{s}')\geq \frac{\delta}{2})&=1-P(\bar{C}(\mathcal{J}_t,\mathbf{s}^*)-\bar{C}(\mathcal{J}_t,\mathbf{s}')\leq \frac{\delta}{2}),
\end{align}
which for $\delta>2L(\frac{2}{M^2})^{\frac{\alpha}{2}}$ converges in probability to $0$ because
\begin{align}
1-P(\bar{C}(\mathcal{J}_t,\mathbf{s}^*)-\bar{C}(\mathcal{J}_t,\mathbf{s}')\leq \frac{\delta}{2})\leq 1-P(\bar{C}(\mathcal{J}_t,\mathbf{s}^*)-\bar{C}(\mathcal{J}_t,\mathbf{s}')\leq L|\mathbf{s}_t-\mathbf{s}'|^{\alpha}\leq \frac{\delta}{2}).
\end{align}
The last equality is a result of the H\"{o}lder continuity properties of the cost function. 
 
Recall that we use the UCB1 algorithm to choose arms within one round of JB. Hence to bound the term  
$P(\bar{C}(\mathcal{J}^*,\mathbf{s}^*)-\bar{C}(\mathcal{J}_t,\mathbf{s}^*)\geq \frac{\delta}{4})$ in \eqref{theorem_one_step_main_eq}, we define two sets of arms (i) set of arms $\mathcal{J}_>$ with sub-optimality gap $\Delta^{\mathcal{J}}_i=\bar{C}(\mathcal{J}^*,\mathbf{s}^*)-\bar{C}(\mathcal{J}^i,\mathbf{s}^*)>\delta/4$ for all $1\leq i \leq N_{mod}$, also referred to as sub-optimal arms and (ii) rest of the arms denoted by ${\mathcal{J}}_<$. Then we have
 \begin{align}\label{eq:one_step_regret_intermediate_step}
P(\bar{C}(\mathcal{J}^*,\mathbf{s}^*)-\bar{C}(\mathcal{J}_t,\mathbf{s}^*)\geq \frac{\delta}{4}) 
=P(\cup_{i\in \boldsymbol{\mathcal{J}}_>}\mathsf{Arm}_t=\mathcal{J}^i)\leq \sum_{i\in \boldsymbol{\mathcal{J}}_>}P(\mathsf{Arm}_t=\mathcal{J}^i),
 \end{align}
where $\mathsf{Arm}_t=\mathcal{J}^i$ indicates that $\mathcal{J}^i$ is chosen at time $t$ and the inequality follows from the union bound. Let $J(T)$ be the (random) set of time steps for which $j_i(T) \leq 8 \log T/ (\Delta^{\mathcal{J}}_i)^2$ for some sub-optimal signaling scheme $i$ in $\{ 1,2,\ldots, T \}$. The time steps in $J(T)$ are the time steps in which there is a high probability that a sub-optimal signaling scheme is selected. In contrast, for time steps in $J(T)^c := \{1,2,\ldots,T\}\backslash J(T)$, all the arms are selected sufficiently many times to have accurate estimates so that the best arm can be correctly identified with a very high probability. Following the analysis of Theorem 1 in \cite{PeterAuer} for the UCB1 algorithm, it can be shown that at time steps $t \in J(T)^c$, the probability of choosing a sub-optimal signaling scheme $i$ is bounded above by $2t^{-4}$. Hence, for $t \in J(T)^c$,  \eqref{eq:one_step_regret_intermediate_step} can be written as
\begin{align}\label{eq:one_step_regret_intermediate_step}
P(\bar{C}(\mathcal{J}^*,\mathbf{s}^*)-\bar{C}(\mathcal{J}_t,\mathbf{s}^*)\geq \frac{\delta}{4}) 
\leq  \sum_{i\in \boldsymbol{\mathcal{J}}_>}2t^{-4} <2N_{mod}t^{-4},
 \end{align}
 Along similar lines, for the third term in \eqref{theorem_one_step_main_eq}, we have
 \begin{align}
P(\bar{C}(\mathcal{J}_t,\mathbf{s}')-\bar{C}(\mathcal{J}_t,\mathbf{s}_t)\geq \frac{\delta}{4})&<2M^2t^{-4},
\end{align}
for all $t\in[1,T]\backslash S(T)$ where $\mathcal{S}_>$ and $S(T)$ are defined along similar lines as $\mathcal{J}_>$ and $J(T)$  but for the $M^2$ discrete arms that correspond to the discretized space of $\mathsf{JNR}$ and $\rho$. 

Overall, by choosing $\delta>2L(\frac{2}{M^2})^{\frac{\alpha}{2}}$ (specifically, we choose $\delta=4L(\frac{2}{M^2})^{\frac{\alpha}{2}}$), the one step regret can be shown to converge in probability to $0$ as $P\left(\bar{C}(\mathcal{J}^*,\mathbf{s}^*)-\bar{C}(\mathcal{J}_t,\mathbf{s}_t)>\delta \right)\leq2(N_{mod}+M^2)t^{-4}\ \forall t
\in[1,T]\backslash \{\mathcal{J}(T)\cup \mathcal{S}(T)\}$
. Since the one-step regret converges in probability to $0$, it implies that the jammer can reach the optimal value of the cost function by using Algorithm~\ref{alg:1}. The value of $\delta$ is governed by the discretization $M$ and the regret incurred by the jammer is mainly due to discretization of the continuum arm space which for larger values of $M$ (possible as $T$ increases as shown in Algorithm~\ref{alg:1}) tightens the confidence bound on the regret.  
 
While the above analysis presented a confidence bound on the one-step regret, it did so by splitting the choice of the signaling scheme and the continuous parameters. We unify these analyses below. The overall sub-optimality gap of the $i$th arm $1\leq i \leq N_{mod}M^2$, denoted by $\{\mathcal{J}^i,\mathbf{s}^i\}$ (recall that $N_{mod}M^2$ arms can be chosen in one round of JB), is defined as $\bar{C}(\mathcal{J}^*,\mathbf{s}^*)-\bar{C}(\mathcal{J}^i,\mathbf{s}^i)$. Let $u_i(t)$ denote the total number of times the $i$th arm has been chosen until time $t$ and $U(T)$ indicate the set of time instants $t\in[1,T]$ for which $u_i(t) \leq \frac{8\log(T)}{\Delta_i^2}$ for some sub-optimal arm $i \in {\cal U}_>$ with a sub-optimality gap $\Delta_i$. It is easy to see that these definitions directly result in the following relationships: $U_{>}\subseteq \mathcal{J}_>\cup \mathcal{S}_>$ and $U(T)\subseteq J(T)\cup S(T)$. 

Note that if the jammer knows a lower bound $\underline{\Delta}>0$ on the minimum sub-optimality gap (for instance, in a wireless communication setting when $SER$ is taken as the cost function, the lower bound can indicate the smallest error in the $SER$ that JB can be allowed to make i.e., we want the choices of JB to be closest to the optimal $SER^*$ that can be achievable) i.e., $\Delta_{\min} := \min_{i \in {\cal U}_>} \Delta_i$, then it can estimate if $t$ belongs to $U(T)$ or not by checking if $u_i(t) > 8 \log T / \underline{\Delta}^2$ for all $i = 1,2,\ldots, N_{mod}$. Let $\hat{U}(T)^c$ be the set of this estimated time slots, for which we have $\hat{U}(T)^c \subseteq U(T)^c$. The jammer will know that for at least $t \in  \hat{U}(T)^c$ the proposed confidence bound will hold true.  While the exact set of all time instants that belong to $U(T)$ may be unknown to the jammer, we can bound the size of the set $E[U(T)]$ as follows \cite{PeterAuer}:
\begin{align}
E[|U(T)|] &\leq \sum_{t=1}^T P(\textrm{a sub-optimal arm } i \in {\cal U}_> \textrm{ is chosen at } t) \notag \\
&\leq 8 \sum_{i \in {\cal U}_>} \left( \frac{\log T}{\Delta_i^2}  \right) + \left(1 + \frac{\pi^2}{3} \right) |U_>|, \notag
\end{align}
which follows from Theorem 1 in \cite{PeterAuer}. This suggests that our confidence bounds hold in all except logarithmically many time slots in expectation.
We stress here that the size of the set $U(T)$ presented above is the worst case bound and the algorithm performs much better than predicted by the bounds as shown in Fig.~\ref{fig:confidence_level} in Section~\ref{Results}. This completes the proof of the Theorem. 


\subsection{Proof of Theorem~\ref{TheoremCumulativeRegret}}
Here, we present a high confidence bound on the cumulative regret incurred by the jammer when it uses Algorithm~\ref{alg:1}. Similar to the regret bound in Theorem~\ref{theorem1}, we have for any given signaling scheme used by the jammer,
{\small \begin{align}\label{risk_bound}
P\left(\sum_{t=1}^T\left(\bar{C}(\mathbf{s}^*)-\bar{C}(\mathbf{s}_t)\right)>\delta\right)&=P\left(\sum_{t=1}^T\left(\bar{C}(\mathbf{s}^*)-\bar{C}(\mathbf{s}')+\bar{C}(\mathbf{s}')-\bar{C}(\mathbf{s}_t)\right)>\delta\right) \nonumber \\
&\hspace{-95pt}\leq P\left(\sum_{t=1}^T\left(\bar{C}(\mathbf{s}^*)-\bar{C}(\mathbf{s}')\right)>\delta/2\right)+P\left(\sum_{t=1}^T\left(\bar{C}(\mathbf{s}')-\bar{C}(\mathbf{s}_t)\right)>\delta/2\right),
\end{align}}
where $\mathbf{s}'$ was defined earlier. For the first term, choosing $\delta>2TL\left(\frac{2}{M^2}\right)^{\alpha/2}=O(T^{\frac{1+2\alpha}{2(1+\alpha)}}(\mathrm{log}T)^{\frac{1}{2(1+\alpha)}})$ and using the H\"{o}lder continuity properties of the cost function gives the following,
{\small \begin{align}\label{risk_bound_term_1}
P\left(\sum_{t=1}^T\left(\bar{C}(\mathbf{s}^*)-\bar{C}(\mathbf{s}')\right)>\delta/2\right)&=1-\left[P\left(\sum_{t=1}^T\left(\bar{C}(\mathbf{s}^*)-\bar{C}(\mathbf{s}')\right)<\delta/2\right)\right] \nonumber \\
&\hspace{-20pt}\leq 1-\left[P\left(\sum_{t=1}^T\left(\bar{C}(\mathbf{s}^*)-\bar{C}(\mathbf{s}')\right)<LT|\mathbf{s^*}-\mathbf{s}'|^{\alpha}<\delta/2\right)\right] {=}0, \ \forall \epsilon >0,
\end{align}}
where the last equality is a result of the choice of $\delta$.

For the second term in \eqref{risk_bound}, as earlier, we use tricks from the UCB1 algorithm to obtain confidence bounds as it is a standard finite-armed $(M^2$ arms$)$ bandit problem \cite{Audibert}. Since the regret in the case of a finite-armed bandit problem can be represented in terms of the number of times an arm has been chosen, we first present a lemma that provides confidence bounds on the same i.e., we evaluate $P(s_k(T)>u)$ where $s_k(T)$ denotes the total number of times arm $k (1\leq k\leq M^2)$ has been chosen until time $T$. Consider the event $\mathcal{E}$ defined by 
\begin{align}\label{cum_regret_step_1}
\hspace{-10pt}\mathcal{E}=\left\{\left\{\forall t: u+1\leq t\leq T \ \mathrm{s.t.}\ B_{k,u,t}\leq \tau \right\}\ \wedge \ \left\{\forall v : 1\leq v\leq T-u \ \mathrm{s.t.}\ B_{k^*,v,u+v}>\tau\right\}\right\},
\end{align}
where $B_{k,u,t}$ indicates the UCB1 index \cite{PeterAuer} of arm $k$ at time instant $t$ when it has been pulled $u$ times until the time time instant $t$, $B_{k^*,v,u+v}$ indicates the UCB1 index of the optimal arm defined along similar lines and $\tau\in \mathbb{R}$ is any real number. In the definition of event $\mathcal{E}$, we have two sub-events, where the first one suggests that the UCB1 index of the $k$th arm is less than $\tau$ and the second event indicates that since arm $k$ was chosen $u$ times, the optimal arm will be chosen $v\in[1,T-u]$ times when its index exceeds $\tau$. Then, for any $1\leq v\leq T-u$ and $u+v\leq t\leq T$ we have $B_{k^*,v,t}\geq B_{k^*,v,u+v}>\tau \geq B_{k,u,t}$ (because we want to compare the indices of the arms at time $t$ given the fact that arm $k$ was chosen $u$ times before this and the optimal arm was chosen $v$ times before this time instant $t$), which suggests that arm $k$ would not be chosen the $(u+1)$th time at any time $t\leq T$ (notice that the exact time instants at which the arms were chosen do not matter, only the number of times an arm chosen decides the UCB1 index of that arm). Thus we have by contradiction
\begin{align}\label{lemma_Tk}
P(s_k(T)>u)&\leq P\left(\exists t:\ u+1\leq t\leq T\ \mathrm{s.t.}\ B_{k,u,t}>\tau \right)\nonumber \\
&\hspace{20pt}+P\left(\exists v:\ 1\leq v\leq T-u\ \mathrm{s.t.}\ B_{k^*,v,u+v}\leq \tau\right).
\end{align}

We will now bound the second term in \eqref{risk_bound} i.e., $P\left(\sum_{t=1}^T\left(\bar{C}(\mathbf{s}')-\bar{C}(\mathbf{s}_t)\right)>\delta/2\right)$. As mentioned earlier, since this is a standard MAB problem, we have the following
{\small \begin{align}
P\left(\sum_{t=1}^T\left(\bar{C}(\mathbf{s}')-\bar{C}(\mathbf{s}_t)\right)>\delta/2\right)&=P\left(\sum_{k:\Delta^{\mathcal{S}}_k>0}\Delta^{\mathcal{S}}_ks_k(T)>\delta/2\right)\leq \sum_{k:\Delta^{\mathcal{S}}_k>0}P\left(s_k(T)>\frac{\delta}{2\Delta^{\mathcal{S}}_k}\right) \nonumber \\
&\leq \sum_{k:\Delta^{\mathcal{S}}_k>0}P\left(s_k(T)>\frac{\delta}{2M^2\Delta^{\mathcal{S}}_k}\right),
\end{align}}
where $\Delta^{\mathcal{S}}_k=\bar{C}(\mathbf{s}')-\bar{C}(\mathbf{s}^{k})$ i.e., the regret incurred for playing arm $k\in[1,M^2]$. 
For any $\delta\geq 16M^2\frac{\mathrm{log}(T)}{\Delta^{\mathcal{S}}_{\min}}$ where $\Delta^{\mathcal{S}}_{\min}$ is the minimum regret incurred across all arms, 
we have $\frac{\delta}{2M^2\Delta^{\mathcal{S}}_k}\geq\frac{8}{(\Delta^{\mathcal{S}}_k)^2}\mathrm{log}(T)$. 
Let $u_k=\ceil{\frac{\delta}{2M^2\Delta^{\mathcal{S}}_k}}$. Using the bound on $s_k(T)$ in \eqref{lemma_Tk} and with $u=u_k$, $\tau=\bar{\mathbf{s}'}$, we have
{\small \begin{align}\label{Tk_bound}
P(s_k(T)>u_k)\leq \sum_{t=u_k+1}^{T}P(B_{k,u_k,t}>\bar{C}(\mathbf{s}'))+\sum_{v=1}^{T-u_k}P(B_{k^*,v,u_k+v}\leq \bar{C}(\mathbf{s}')).
\end{align}}
Using the condition on $\delta$, we have $\sqrt{\frac{2\mathrm{log}(T)}{u_k}}\leq \Delta^{\mathcal{S}}_k/2$ which upon rearranging the terms also gives $T\leq e^{\frac{u_k(\Delta^{\mathcal{S}}_k)^2}{8}}$. Then we can bound $P(B_{k,u_k,t}>\bar{C}(\mathbf{s}'))$  (which indicates the fact that the upper confidence bound on the sub-optimal arm is higher than the mean reward/cost function incurred by the optimal strategy $\mathbf{s}'$) as
{\small \begin{align}\label{cum_reg_inter_eq}
P(B_{k,u_k,t}>\bar{C}(\mathbf{s}'))&\leq P(B_{k,u_k,T}>\bar{C}(\mathbf{s}'))=P\left(\hat{C}(\mathbf{s}^{k})+\sqrt{\frac{2\mathrm{log}(T)}{u_k}}>\bar{C}(\mathbf{s}^k)+\Delta^{\mathcal{S}}_k\right) \nonumber \\
&\leq P\left(\bar{C}(\mathbf{s}^{k})-\bar{C}(\mathbf{s}')>\Delta^{\mathcal{S}}_k/2\right) \leq e^{-u_k(\Delta^{\mathcal{S}}_k)^2/2},
\end{align}}
where $\hat{C}(\mathbf{s}^{k})$ indicates the estimate of the cost function/reward obtained by suing strategy $\mathbf{s}^k$, the second inequality is obtained by using the definition of $u_k$ and $\delta$, and the third inequality by using the Chernoff-Hoeffding bound. Using \eqref{cum_reg_inter_eq} and the fact that $T\leq e^{\frac{u_k(\Delta^{\mathcal{S}}_k)^2}{8}}$, the first summation in \eqref{Tk_bound} can be upper bounded by $e^{-\frac{u_k(\Delta^{\mathcal{S}}_k)^2}{4}}$.

The second summation in \eqref{Tk_bound} is bounded by using the Chernoff-Hoeffding bound as\newline
$\sum_{v=1}^{T-u_k}P(B_{k^*,v,u_k+v}\leq \bar{C}(\mathbf{s}'))\leq \sum_{v=1}^{T-u_k}(u_k+v)^{-4}\leq \sum_{y=u_k}^{T}y^{-4}\leq \int_{y=u_k}^{\infty}y^{-4}=\frac{u_k^{-3}}{3}$ (by change of variable). Overall, \eqref{Tk_bound} can be upper bounded as  
{\small \begin{align}
P(s_k(T)>u_k)\leq e^{-\frac{u_k(\Delta^{\mathcal{S}}_k)^2}{4}}+\frac{u_k^{-3}}{3}.
\end{align}}
For $u_k=\ceil{\frac{\delta}{2M^2\Delta^{\mathcal{S}}_k}}$, the above bound is given by
{\small \begin{align}
P\left(s_k(T)>\frac{\delta}{2M^2\Delta^{\mathcal{S}}_k}\right)=P\left(s_k(T)>\ceil{\frac{\delta}{2M^2\Delta^{\mathcal{S}}_k}}\right) \leq e^{-\frac{\delta\Delta^{\mathcal{S}}_k}{8M^2}}+8\delta^{-3}\frac{(M^2\Delta^{\mathcal{S}}_k)^{3}}{3}.
\end{align}}
Thus we have the upper bound on the second term in \eqref{risk_bound} as 
{\small \begin{align}\label{second_term_cum_regret}
P\left(\sum_{t=1}^T\left(\bar{C}(\mathbf{s}')-\bar{C}(\mathbf{s}_t)\right)>\delta/2\right)&\leq \sum_{k:\Delta_k>0}\left\{e^{-\frac{\delta\Delta_k}{8M^2}}+8\delta^{-3}\frac{(M^2\Delta^{\mathcal{S}}_k)^{3}}{3}\right\} \nonumber \\
&\leq (M^2-1)\left(1+\frac{8}{3}\delta^{-3}M^6\right)
\approx \frac{8}{3}\delta^{-3}M^8\triangleq \epsilon,
\end{align}}
where we used the fact that $\Delta_k\in[0,1]$. Overall using \eqref{risk_bound_term_1} and \eqref{second_term_cum_regret}, for all \break $\delta>\max\left(2TL\left(\frac{2}{M^2}\right)^{\alpha/2},16M^2\frac{\mathrm{log}(T)}{\Delta^{\mathcal{S}}_{\min}}\right)$, we have 
{\small \begin{align}\label{cum_regret_bound}
P\Big(\sum_{t=1}^T\Big(\bar{C}(\mathbf{s}^*)-\bar{C}(\mathbf{s}_t)\Big)>\left(\frac{8}{3\epsilon}\Big(\frac{T}{\mathrm{log}(T)}\right)^{\frac{4}{1+\alpha}}\Big)^{1/3}\Big)<\epsilon.
\end{align}}
Since $\Delta_{\min}$ is unknown \emph{a priori}, a jammer can use any known lower bound $\underline{\Delta}$ to obtain $\delta$. This lower bound can be obtained as described in the proof of Theorem 3.

Similar bounds i.e., of the same order can be obtained irrespective of the signaling scheme used by the jammer (here the jammer uses $N_{mod}$ signaling schemes) as it is just another finite armed bandit problem. 
This can be done as follows
\begin{align}
P\Big(\sum_{t=1}^T(\bar{C}(\mathcal{J}^*,\mathbf{s}^*)&-\bar{C}(\mathcal{J}_t,\mathbf{s}_t)>\delta)\Big)=
P\Big(\sum_{t=1}^T(\bar{C}(\mathcal{J}^*,\mathbf{s}^*)-\bar{C}(\mathcal{J}^*,\mathbf{s}_t)+\bar{C}(\mathcal{J}^*,\mathbf{s}_t)-\bar{C}(\mathcal{J}_t,\mathbf{s}_t)>\delta)\Big)\nonumber \\
&\hspace{-40pt}\leq P\Big(\sum_{t=1}^T(\bar{C}(\mathcal{J}^*,\mathbf{s}^*)-\bar{C}(\mathcal{J}^*,\mathbf{s}_t)>\delta/2)\Big)+
P\Big(\sum_{t=1}^T(\bar{C}(\mathcal{J}^*,\mathbf{s}_t)-\bar{C}(\mathcal{J}_t,\mathbf{s}_t)>\delta/2)\Big),
\end{align}
where the first term in the above equation can be directly bounded by using the analysis presented for \eqref{cum_regret_bound} and the second term can be bounded directly using the properties of the UCB1 algorithm since it is a discrete arm selection (specifically, by using the steps in \eqref{cum_regret_step_1}-\eqref{cum_regret_bound}).

\section*{Appendix D\\ High Confidence Bound on Estimates}
{\emph Proof of Theorem~\ref{TheoremEstimateProb}:}
A high confidence bound on the mean estimate of the reward/cost function for any strategy that is used at time $t$ by the jammer is developed. To do so, we evaluate $P(\bar{C}(\mathcal{J}^*,\mathbf{s}^*)-\hat{C}(\mathcal{J}_t,\mathbf{s}_t)>\delta)$ as follows,
{\small \begin{align}\label{estimate_prob_bound}
P(\bar{C}(\mathcal{J}^*,\mathbf{s}^*)-\hat{C}(\mathcal{J}_t,\mathbf{s}_t)>\delta)\leq P(\bar{C}(\mathcal{J}^*,\mathbf{s}^*)-\bar{C}(\mathcal{J}_t,\mathbf{s}_t)>\frac{\delta}{2})+P(\bar{C}(\mathcal{J}_t,\mathbf{s}_t)-\hat{C}(\mathcal{J}_t,\mathbf{s}_t)>\frac{\delta}{2}),
\end{align}}
\noindent where $\bar{C}(\mathcal{J}_t,\mathbf{s}_t)$ is the actual mean reward/cost of the strategy $(\mathcal{J}_t,\mathbf{s}_t)$.
The first term can be bounded using Theorem~\ref{TheoremOneStepRegret} where it can be shown to be less than $2(N_{mod}+M^2)t^{-4}$ for all $\delta>2^{\frac{5\alpha+4}{2(1+\alpha)}}L^{\frac{1}{1+\alpha}}\left(\frac{\mathrm{log}T}{T}\right)^{\frac{\alpha}{2(1+\alpha)}}$ and $t\in[1,T]\backslash \{U(T)\}$, where $U(T)$ is defined in the proof of Theorem~\ref{TheoremOneStepRegret}. 
For the second term, notice that we are comparing the actual and estimated mean rewards of the strategy $(\mathcal{J}_t,\mathbf{s}_t)$ which can be bounded using the Chernoff-Hoeffding bound and the properties of the UCB1 algorithm as follows, 
\begin{align}
P(\bar{C}(\mathcal{J}_t,\mathbf{s}_t)-\hat{C}(\mathcal{J}_t,\mathbf{s}_t)>\frac{\delta}{2})\leq \exp(-\frac{u(t)\delta^2}{2}),
\end{align}
where $u(t)$ is the total number of times the strategy $(\mathcal{J}_t,\mathbf{s}_t)$ has been used until time $t$. Since we use the UCB1 algorithm, in all the time instants $t$ in which the arm $(\mathcal{J}_t,\mathbf{s}_t)$ has been chosen at least $\frac{8\mathrm{log}t}{\Delta_t^2}$ where ($\Delta_t=\bar{C}(\mathcal{J}^*,\mathbf{s}^*)-\bar{C}(\mathcal{J}_t,\mathbf{s}_t)$ is the sub-optimality gap of the strategy $(\mathcal{J}_t,\mathbf{s}_t)$), we have $P(\bar{C}(\mathcal{J}_t,\mathbf{s}_t)-\hat{C}(\mathcal{J}_t,\mathbf{s}_t)>\frac{\delta}{2})\leq \exp(-16\mathrm{log}t)=t^{-16}$, because $\Delta_t\leq \delta/2$ from the bound on the first term. This completes the proof of the Theorem. 


\section*{Appendix E \\ Jamming Bandits with Arm Elimination}
Algorithm~\ref{alg:2} is the modified JB algorithm that uses the UCB-Improved algorithm \cite{AuerElimination} instead of the UCB1 algorithm. Even in this algorithm, the regret has two terms, similar to the earlier regret bounds presented in \eqref{regret_term_1} and \eqref{regret_term_2}. While the bound in \eqref{regret_term_1} continues to hold true even in this case, the bound in \eqref{regret_term_2} changes due to the UCB-Improved algorithm. From \cite{AuerElimination}, the regret bound for a $M^2$-armed bandit algorithm is given by $\mathcal{O}(\sqrt{M^2T}\frac{\mathrm{log}(M^2\mathrm{log}(M^2))}{\sqrt{\mathrm{log}(M^2)}})$. Thereby the overall regret is $TL\left(\frac{2}{M^2}\right)^{\frac{\alpha}{2}}+O\left(\sqrt{M^2T}\frac{\mathrm{log}(M^2\mathrm{log}(M^2))}{\sqrt{\mathrm{log}(M^2)}}\right)$. By using the doubling trick, the regret bounds for the overall time horizon $n$ can be obtained.

\section*{Appendix F \\ Jamming Bandits against Time-Varying Victims}
For a given signaling scheme $\mathcal{J}$ used by the jammer, the average regret (i.e., with respect to the  victim's strategies) can be bounded as below,
\begin{align}
R_n&=\Big[\sum_{t=1}^n\sum_{i=1}^{|\mathcal{P}|}p_i\Big(\mathbf{E}(C^{i}_t(\mathbf{s}^*))-\mathbf{E}(C^{i}_t(\mathbf{s}_t))\Big)\Big] =\Big[\sum_{t=1}^n\sum_{i=1}^{|\mathcal{P}|}p_i\Big(\bar{C}^i_t(\mathbf{s}^*)-\bar{C}^i_t(\mathbf{s}_t)\Big)\Big] \nonumber \\
&=\Big[\sum_{t=1}^n\sum_{i=1}^{|\mathcal{P}|}p_i\Big(\{\bar{C}^i_t(\mathbf{s}^*)-\bar{C}^i_t(\mathbf{s}')\}+\{\bar{C}^i_t(\mathbf{s}')-\bar{C}^i_t(\mathbf{s}_t)\}\Big)\Big],
\end{align}
where $\bar{C}$ indicates the average cost function, $\mathbf{s}^*$ is the optimal jamming strategy across all stochastic strategies that can be used by the victim (since the jammer is not aware of the action taken by the user), $\mathbf{s}'$ is the strategy closest in Euclidean distance to the optimal strategy (as defined in the Section~\ref{sec:FixedUser}) and $\mathbf{s}_t$ is the actual strategy chosen 
at time $t$. The first term can be bounded using the H\"{o}lder continuity properties of the cost function (see Appendix B where a similar analysis is done for the fixed user strategy case). The second term is bounded as follows. Since the jammer is not aware of the users' strategy, partition of the action space is done a priori and hence we use the same discretization $M$ across all users' strategies. Thus define $\Delta=\sqrt{M^2\mathrm{log}(T)/T}$. For each action $i$ taken by the user, define $\Delta_{ij}=\bar{C}(s')-\bar{C}(s^j)$ as the loss in rewards when a sub-optimal arm $j$ is chosen. We split the arms into two sets, a) for those arms which satisfy $\Delta_{ij}<\Delta$, the near optimal arms and b)  $\Delta_{ij}>\Delta$, the sub-optimal arms. For the first set, the maximum regret incurred against each user strategy is $T\Delta$ (over a time period $T$ of the inner loop in Algorithm~\ref{alg:1}). For the second set, since we use the UCB1 algorithm, each sub-optimal arm is only chosen a maximum of $\frac{8\mathrm{log}(T)}{\Delta_{ij}^2}+1+\frac{\Pi^2}{3}$ times \cite{PeterAuer}. Thus we can bound the regret for the arms in the second set as $\Delta_{ij}\left(\frac{8\mathrm{log}(T)}{\Delta_{ij}^2}+1+\frac{\Pi^2}{3}\right)$. Overall the regret is upper bounded by
$\sum_{i=1}^{|\mathcal{P}|}p_i\Big(TL\left(\frac{2}{M^2}\right)^{\alpha/2}+T\Delta+\sum_{j=1}^{M^2}\Delta_{ij}\left[\frac{8\mathrm{log}(T)}{\Delta_{ij}^2}+1+\frac{\Pi^2}{3}\right]\Big)$,
 which by using the relationship between $\Delta$ and $\Delta_{ij}$ and the fact that $\Delta_{ij}\in[0,1]$ can further be upper bounded as
\begin{align}\label{distribution_regret}
\Big(TL\left(\frac{2}{M^2}\right)^{\alpha/2}+T\Delta+M^2\Big[\frac{8\mathrm{log}(T)}{\Delta}+1+\frac{\Pi^2}{3}\Big]\Big)\approx \mathcal{O}(T^{\frac{\alpha+2}{2(\alpha+1)}} \mathrm{log}^{\frac{\alpha}{2(\alpha+1)}} T).
\end{align}
Since the jammer can choose from $N_{mod}$ signaling schemes, the overall regret is the scaled (by $N_{mod}$) version of \eqref{distribution_regret}, which is similar to the regret bound in Theorem~\ref{theorem1}. Remember this is just a upper bound on the regret and that the actual regret depends on the strategy employed by the victim transmit-receive pair. Also notice that, all the regret and high confidence bounds derived in Section~\ref{sec:FixedUser} can also be extended to this scenario as well.
\end{document}